\documentclass[a4paper,11pt]{article}
\pdfoutput=1 

\usepackage{jheppub} 
\usepackage[english]{babel}
\usepackage[utf8x]{inputenc}
\usepackage[T1]{fontenc}
\usepackage{epstopdf}
\usepackage{float}
\usepackage{subcaption}
\usepackage{amssymb}
\usepackage[T1]{fontenc} 
\usepackage{amsmath}
\usepackage{graphicx}
\usepackage[colorinlistoftodos]{todonotes}
\usepackage{float}

\title{\boldmath Tunable Quantum Chaos in the Sachdev-Ye-Kitaev Model Coupled to a Thermal Bath}

\author[a,b]{Yiming Chen}
\author[a,c]{Hui Zhai}
\author[a]{Pengfei Zhang\note{E-mail: PengfeiZhang.physics@gmail.com}}

\affiliation[a]{Institute for Advanced Study, Tsinghua University, 100084, Beijing, China}
\affiliation[b]{Department of Physics, Tsinghua University, 100084, Beijing, China}
\affiliation[c]{Collaborative Innovation Center of Quantum Matter, Beijing, 100084, China}

\abstract{The Sachdev-Ye-Kitaev (SYK) model describes Majorana fermions with random interaction, which displays many interesting properties such as non-Fermi liquid behavior, quantum chaos, emergent conformal symmetry and holographic duality. Here we consider a SYK model or a chain of SYK models with $N$ Majorana fermion modes coupled to another SYK model with $N^2$ Majorana fermion modes, in which the latter has many more degrees of freedom and plays the role as a thermal bath. For a single SYK model coupled to the thermal bath, we show that although the Lyapunov exponent is still proportional to temperature, it monotonically decreases from $2\pi/\beta$ ($\beta=1/(k_BT)$, $T$ is temperature) to zero as the coupling strength to the thermal bath increases. For a chain of SYK models, when they are uniformly coupled to the thermal bath, we show that the butterfly velocity displays a crossover from a $\sqrt{T}$-dependence at relatively high temperature to a linear $T$-dependence at low temperature, with the crossover temperature also controlled by the coupling strength to the thermal bath. If only the end of the SYK chain is coupled to the thermal bath, the model can introduce a spatial dependence of both the Lyapunov exponent and the butterfly velocity. Our models provide canonical examples for the study of thermalization within chaotic models.}

\begin{document} 
\maketitle
\unitlength = 1mm
\flushbottom

\section{Introduction}

As a concrete solvable model displaying maximal quantum chaos, non-Fermi liquid behavior and holographic duality, the Sachdev-Ye-Kitaev (SYK) model proposed by Kitaev \cite{Kitaev2} based on the early work of Sachdev and Ye \cite{SY} has been widely studied recently both on the field-theory side \cite{Kitaev2,Comments,spectrum1,spectrum2,spectrum3,Liouville,Liouville2,SYK new,SYK new2,SYK new3} and the gravitational side \cite{Comments,bulk Yang,bulk spectrum Polchinski,bulk2,bulk3,bulk4,bulk5,syk-bh,SYK g new1,SYK g new2,SYK g new3}. On the field theory side, the model contains $N$ Majorana fermions $\chi_i\ (i=1,...,N)$ with random four-fermion interaction terms. The Hamiltonian is
\begin{equation}
H_{\chi}=\frac{1}{4!}\sum_{ijkl}^{N}J_{ijkl}\chi_i \chi_j \chi_k \chi_l, \label{SYK}
\end{equation}
where the normalization of $\chi_i$ is given by anticommutation relation $\{\chi_i, \chi_j \}=\delta_{ij}$. The coupling constants $\{J_{ijkl}\}$ are antisymmetric with respective exchanging of indices. Their mean vanishes and the variance is finite as
\begin{equation}\label{eq2.2}
\overline{J_{ijkl}}=0,\,\,\,\,\,\,\overline{J_{ijkl}^{2}}=\frac{3!J^2}{N^3}.
\end{equation}
The interaction is all-to-all, thus the model is usually viewed as a (0+1)-d quantum mechanics model. The specific choice of $N$ scaling in Eq. (\ref{eq2.2}) leads to an elegant large-$N$ structure, with which the two-point and four-point correlation functions can be calculated analytically \cite{Comments}. The exact solution in large-$N$ limit displays a number of very intriguing properties, as we briefly summarized below. 

First, it displays an emergent conformal symmetry and the non-Fermi liquid behavior. To the leading order of $\frac{1}{N}$ expansion, the imaginary-time two-point function $G(\tau)\delta_{ij}\equiv\langle\mathcal{T}_{\tau}\chi_i (\tau)\chi_j (0) \rangle$ satisfies a simple form of Schwinger-Dyson equations as
\begin{equation}\label{eq2.5}
G(i\omega)^{-1}=-i\omega -\Sigma(i\omega),\,\,\, \Sigma(\tau)=J^2 G(\tau)^3.
\end{equation}
In the infrared limit, one drops the $-i\omega$ term, and the Schwinger-Dyson equations can be solved by using following ansatz
\begin{equation}
G(\tau)=b\frac{\textrm{sgn}(\tau)}{|\tau|^{\frac{1}{2}}},
\end{equation}
which leads to 
\begin{equation}
J^2b^4=\frac{1}{4\pi}.
\end{equation}
A Fourier transformation of the Green's function shows the divergence of spectral function as $1/\omega^{1/2}$ at low energy, which signals a non-Fermi Liquid behavior. The form of the Green's function also indicates that the system acquires an emergent conformal symmetry in the IR limit with the scaling dimension of $\chi$ equals to $1/4$, which can also be seen from the fact that the fix-point action only has the four-fermion interaction terms. With the help of this conformal symmetry and by utilizing conformal mapping $\tau\rightarrow \tan \frac{\pi \tau}{\beta} $, the low temperature Green's function (with $\beta J\gg1$) is then given by
\begin{equation}
G_{\beta}(\tau)=b\frac{\textrm{sgn}(\tau)}{\left|\frac{\beta}{\pi}\sin\left(\frac{\pi\tau}{\beta}\right)\right|^{\frac{1}{2}}}.
\end{equation}

Second, it shows maximally chaotic behavior. With the knowledge of the two-point Green's function at finite temperature, one can further calculate the four-point Green's function of the SYK model within the large-$N$ expansion. At the leading order $\mathcal{O}(N^0)$, the four-point function is given by its disconnected part, as the connected part starts at order $\mathcal{O}(\frac{1}{N})$
\begin{equation}
\frac{1}{N}\mathcal{F}(\tau_1,\tau_2,\tau_3,\tau_4)=\frac{1}{N^2}\sum_{j,k=1}^{N}\langle \mathcal{T}_{\tau}\chi_j (\tau_1) \chi_j (\tau_2)\chi_k (\tau_3)\chi_k (\tau_4)\rangle - G(\tau_{12})G(\tau_{34}).
\end{equation}
With the analytical continuation of $\mathcal{F}(\tau_1,\tau_2,\tau_3,\tau_4)$ to $\mathcal{F}(\frac{3\beta}{4}+it,\frac{\beta}{4}+it,\frac{\beta}{2},0)$, one can calculate the out-of-time-ordered correlation (OTOC) function for $\chi_i$ and $\chi_j$. The definition for a regularized OTOC for operator $A$ and $B$ is:
\begin{equation}
F_{AB}(t)=\text{Tr}[yA^\dagger(t)yB^\dagger(0)yA(t)yB(0)],~~~~ \textrm{where}~~y=e^{-\frac{1}{4}\beta\hat{H}}.
\end{equation}
which diagnoses the quantum butterfly effect \cite{Kitaev1,bh1,bh2,bh3} and has been applied to various models recently \cite{otoc cft,otoc anyon,OTOC-Boson Hubbard,OTOC-Keldysh,OTOC-Luttinger,OTOC-MBL2,OTOC-MBL3,OTOC-MBL4,OTOC-MBL5,OTOC-MBL6,OTOC-Ruihua MBL1,OTOC-Yao,OTOC-classical,OTOC-quantum,OTOC-quantum channel,OTOC-protocal,OTOC new}. There are also experimental measurements of OTOC in different systems \cite{OTOC-exp1,OTOC-exp2}. The OTOC in many chaotic systems has a universal behavior as $F_{AB}(t)\sim c_0-c_1\exp(\lambda_Lt)$ at a time scale called the scrambling time $t_s$. $\lambda_L$ defines the Lyapunov exponent for a quantum system. Assuming $t_s\gg t_d$ ($t_d$ is defined as the decay time for two point function), it is proved that $\lambda_L$ is bounded by $2\pi/\beta$, and the models with holographic duality is believed to saturate the bound \cite{Kitaev2,bh1,bh2,bh3,prove}. For the SYK model, one can explicitly show that $F(t)\equiv F_{\chi_i\chi_j}(t)\sim -\beta J\exp(2\pi t/\beta)$ when $t\gg t_d=\beta$. The Lyapunov exponent extracted from it is exactly $2\pi/\beta$.

Thirdly, a gravitational model shares the same effective action as the low-energy theory of the SYK model. In the SYK model, by using the replica trick to treat the disorder average, a non-local action can be deduced after introducing bi-local fields $G(\tau_1,\tau_2)$ and $\Sigma (\tau_1,\tau_2)$:
\begin{equation}
\frac{S}{N}=-\frac{1}{2}\log\det (\partial_\tau - \Sigma)+\frac{1}{2}\int d\tau_1 d\tau_2 \left[\Sigma (\tau_1,\tau_2)G(\tau_1,\tau_2) - \frac{J^2}{4}G(\tau_1,\tau_2)^4 \right].
\end{equation}
The saddle point equations for $G(\tau_1,\tau_2)$ and $\Sigma (\tau_1,\tau_2)$ (by assuming the translational invariance in time) are the same as shown in Eq. (\ref{eq2.5}). Using saddle-point solutions, one can show there is a finite zero-temperature entropy which is another signature of a non-Fermi Liquid resulted from the large degeneracy of ground states \cite{numerics wenbo}. Expanding $G$ and $\Sigma$ around their saddle point solutions gives the low-energy effective action for the SYK model. Note that the low energy physics of the SYK model is dominated by the reparametrization modes due to the emergent conformal symmetry, and a reparametrization to the time variable $\tau \rightarrow f(\tau)$ acts on the $G(\tau_1,\tau_2)$ field as $G(\tau_1,\tau_2)\rightarrow (f'(\tau_1)f'(\tau_2))^{\frac{1}{4}}G(f(\tau_1),f(\tau_2))$. For small deformation $\tau\rightarrow \tau + \epsilon (\tau)$, the effective action can be approximated by an action of the fluctuation $\delta_{\epsilon}G$, which has an elegant form as the Schwarzian action:
\begin{equation}
\frac{S}{N}\propto \int_{0}^{\beta}d\tau \frac{1}{2}\left[(\epsilon'')^2 - \left(\frac{2\pi}{\beta}\right)^2 (\epsilon')^2\right].
\end{equation}
For finite transformation $\tau \rightarrow f(\tau)$, the action can be written as
\begin{equation}
\frac{S}{N}= -\frac{\#}{J}\int d\tau \left\{\tan\left(\frac{\pi f}{\beta}\right),\tau\right\},\,\,\,\, \textrm{where}\,\, \{f,\tau\}\equiv \frac{f'''}{f'}-\frac{3}{2}\left(\frac{f''}{f'}\right)^2.
\end{equation}
$\#$ is some constant number given in Ref.\cite{Comments}. This Schwarzian effective action roots in the conformal symmetry of the SYK model and can give rise to the maximal Lyapunov exponent. Interestingly, on the gravity side, the same action appears for the dilaton gravity theory in two-dimensional near anti-de Sitter spacetime (NAdS$_2$) \cite{bulk Yang}.

Motivated by the aforementioned intriguing properties of the SYK model, recently there appear many interesting generalizations of the SYK model \cite{numerics wenbo,Yingfei1,Yingfei2,generalization 1,wenbo susy,susy2,no disorder1,no disorder2,no disorder3,no disorder4,no disorder5,no disorder6,no disorder7,no disorder8,no disorder9,no disorder10,Altman,thermal transport,high-D1,high-D2-con,yyz condensation,sk jian,generalization 2,transition1,our,Balent,thick,new gsyk1,no disorder new,susy3}. The models with $U(1)$ symmetry are studied numerically for both bosons and fermions in Ref.\cite{numerics wenbo} by exact diagonalization, which shows the nearly degeneracy of the many-body spectrum at low energy for the fermionic model and the spin-glass state for the bosonic model. There are also efforts toward a more precise holographic description for which  a supersymmetric version of the SYK model is studied \cite{wenbo susy,susy2,susy3}. By coupling the SYK models  \cite{Yingfei1,Yingfei2}, an lattice model with spatial degree of freedom can be realized, where the butterfly velocity and the diffusion constant can be calculated and are compared with the holographic result. The issue of chaotic to non-chaotic transiton is considered with several different generalization of the SYK models \cite{Altman,sk jian,yyz condensation,our,Balent}. Among all these generalizations, the non-Fermi Liquid phases given by the SYK-type models all possess a maximal chaotic behavior and Lyapunov exponent saturating to $2\pi/\beta$, with only one exception that contains time dependent interaction \cite{Comments,thick}.

In this paper, we consider a SYK model or a chain of SYK models with $N$ Majorana fermion modes coupled to another SYK model with $N^2$ Majorana fermion modes. The latter contains many more degrees of freedom and can be viewed as as a kind of thermal bath. Our model is time independent and this model is also solvable in the large-$N$ limit. In view of the three properties mentioned above, we will show that, on one hand, this model still displays an emergent conformal symmetry and a non-Fermi liquid behavior, but on the other hand, it is {\textit{not}} maximally chaotic. In fact, we will show that the Lyapunov exponent of this model can be tuned to any value between zero and $2\pi/\beta$. Whether this model also has a gravitational dual remains unclear.    

In section 2, we discuss a single SYK model coupled to the thermal bath. We show analytically illustrate that $\lambda_L $ for the small system monotonically decreases from $2\pi /\beta$ to zero as the coupling strength to the thermal bath increases. 
In section 3, we consider a chain of SYK models. When the chain is uniformly coupled to the thermal bath, the butterfly velocity displays a crossover from $\sqrt{T}$-dependence at relatively high temperature to a new linear $T$-dependence at low temperature. If only the end of the SYK chain is coupled to the thermal bath, we found that there is a spatial dependence of both the Lyapunov exponent and the butterfly velocity.

\section{$(0+1)$-d SYK Model Coupled to a Thermal Bath}

%

\subsection{The Model and the Two-Point Functions}

In this section, we introduce a generalized version of the SYK model that contains a small SYK model with $N$ Majorana fermions (denoted by $\chi_i$, $i=1,2,...,N$) to a large SYK model with $N^2$ Majorana fermions (denoted by $\psi_i$, $i=1,2,...,N^2$). We will call the two subsystem as SYK$_{\chi}$ and SYK$_{\psi}$, respectively. In fact, the number of Majorana fermions in the large SYK model does not have to be $N^2$, and it can be generally $N^\alpha$ as long as $\alpha>1$ such that it dominates over $N$ in the large $N$ limit. Here we choose the number of Majorana fermions in the larger cluster to be $N^2$ just for concreteness. The Hamiltonian of the coupled SYK system is written as:
\begin{equation}
H =H_{\chi}+H_{\psi}+H_{c}\,\,,
\end{equation}
where both $H_{\chi}$ and $H_{\psi}$ take the same form as Eq. (\ref{SYK}), with $\{J_{ijkl}\},\{J^{'}_{ijkl}\}$ being the random couplings in $H_{\chi}$ and $H_{\psi}$, respectively. $H_{c}$ is the coupling Hamiltonian defined as
\begin{equation}\label{notation}
H_{c}=\frac{1}{4} \sum_{ijkl}u_{ijkl}\chi_{i}\chi_{j}\psi_{k}\psi_{l},
\end{equation}
where $u_{ijkl}$ are also random coupling numbers. $\{J_{ijkl}\},\{J^{'}_{ijkl}\}, \{u_{ijkl}\}$ are antisymmetric random variables with zero mean
\begin{equation}
\overline{J_{ijkl}}=0,\,\,\,\,\,\,\overline{J_{ijkl}^{'}}=0,\,\,\,\,\,\, \overline{u_{ijkl}}=0,
\end{equation}
and their variances are 
\begin{equation}\label{eq3}
\overline{J_{ijkl}^{2}}=\frac{3!J^2}{N^3},\,\,\,\,\,\, \overline{J_{ijkl}^{'2}}=\frac{3!J^2}{N^6},\,\,\,\,\,\,
\overline{u_{ijkl}^{2}}=\frac{2!\, u^2}{N^5}.
\end{equation}
The reason for making this specific choice of $N$ dependence of the variances will be clear soon because it gives a nice structure for the theory in the large-$N$ limit. 

\begin{figure}[t]
	\centering
	\includegraphics[width=0.9\textwidth]{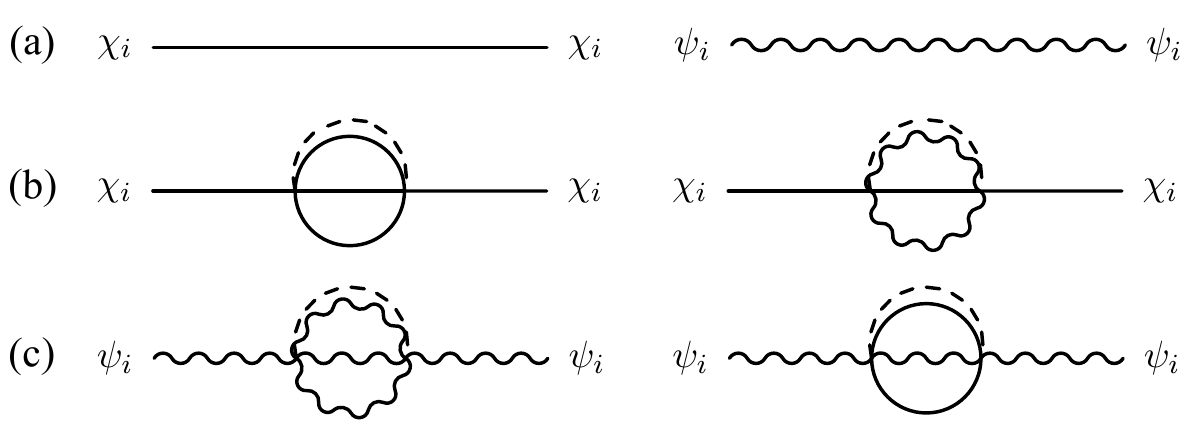}
	\caption{(a). Left: The Green's function $G_{\chi}(\tau)$ of SYK$_{\chi}$; Right: The Green's function $G_{\psi}(\tau)$ of SYK$_{\psi}$. (b). Order of $N$ for self energy diagram of $\chi$: Left: $\sim\frac{J^2}{N^3}\times N^3 \sim \mathcal{O}(N^{0})$; \,\,Right: $\sim \frac{u^2}{N^{5}}\times N\times (N^2)^2 \sim \mathcal{O}(N^{0})$. The dashed line means the two vertices should have same coupling. (c). Order of $N$ for self energy diagram of $\psi$: Left: $\sim\frac{J^2}{N^6}\times (N^2)^3 \sim \mathcal{O}(N^{0})$;\,\, Right: $\sim \frac{u^2}{N^5}\times N^2\times N^2 \sim \mathcal{O}(\frac{1}{N})$.}\label{Feydia1}
\end{figure}

As shown in Fig. \ref{Feydia1} (a), we use straight line and wavy line to denote their Green's functions $G_{\chi}(\tau)=\langle \mathcal{T}_{\tau} \chi_{i}(\tau)\chi_{i}(0)\rangle$ and $G_{\psi}(\tau)=\langle \mathcal{T}_{\tau} \psi_{i}(\tau)\psi_{i}(0)\rangle$, respectively, where $\mathcal{T}_{\tau}$ denotes time-order operator in the imaginary time $\tau$. Note that the Green's functions are diagonal in term of the fermion indices because of the disorder average. In the large-$N$ limit, the Green's functions can be be determined by diagrammatic method and by solving the Schwinger-Dyson equations.
For the SYK$_{\chi}$, one would expect two contributions to its self-energy $\Sigma_{\chi}(\tau)$, as given by two diagrams in Fig. \ref{Feydia1} (b). The choices of the orders of $N$ in Eq. (\ref{eq3}) ensures that the same two diagrams are both of order $\mathcal{O}(N^{0})$, which is the lowest order in $\frac{1}{N}$ for the Green's function. From these diagrams we can obtain the Schwinger-Dyson equations for the SYK$_{\chi}$ system as
\begin{equation}\label{eq4}
G_{\chi}(\omega)^{-1} = -i\omega - \Sigma_{\chi}(\omega), ~~~~ \Sigma_{\chi}(\tau)=J^2 G_{\chi}^{3}(\tau)+u^2 G_{\psi}^{2}(\tau)G_{\chi}(\tau).
\end{equation}

For the SYK$_{\psi}$ system, in the large-$N$ limit, one may also expect two contributions for its self-energy $\Sigma_{\psi}(\tau)$, as shown in Fig. \ref{Feydia1}. However, the choices in Eq. (\ref{eq3}) results in the suppression of the second diagram by $\frac{1}{N}$, thus it can be neglected at the leading order. This large-$N$ structure physically makes sense, since physically the properties of the larger system should not be affected by the small system at the leading order of $\frac{1}{N}$. Then we have the Schwinger-Dyson equations for the SYK$_{\psi}$ system:
\begin{equation}\label{eq5}
G_{\psi}(\omega)^{-1}=-i\omega -\Sigma_{\psi}(\omega),~~~~ \Sigma _{\psi}(\tau)=J^2 G_{\psi}^{3}(\tau).
\end{equation}

In the strong coupling limit, we first drop the $-i\omega$ term, and the Green's functions obey the form as the original single SYK model 
\begin{equation}\label{eq6}
G_{\chi}(\tau)= a\frac{\text{sgn}(\tau)}{|\tau|^{\frac{1}{2}}},~~~~ G_{\psi}(\tau)= b\frac{\text{sgn}(\tau)}{|\tau|^{\frac{1}{2}}},
\end{equation}
where the coefficients $a$ and $b$ are determined by
\begin{equation}\label{eq7}
a^4 J^2 +u^2 a^2 b^2 =\frac{1}{4\pi},~~~~ b^4 J^2 = \frac{1}{4\pi}.
\end{equation}
Note that the coefficients are different from the original SYK model, which is the key to the discussion below. Consequently, the finite temperature version of the Green's functions are
\begin{equation}
G_{\chi}(\tau)= a\left[\frac{\pi}{\beta \sin \frac{\pi \tau}{\beta}}\right]^{\frac{1}{2}}\text{sgn}(\tau),~~~~ G_{\psi}(\tau)= b\left[\frac{\pi}{\beta \sin \frac{\pi \tau}{\beta}}\right]^{\frac{1}{2}}\text{sgn}(\tau).
\end{equation}

\subsection{Four Point Functions and the Tunable Lyapunov Exponent }\label{Tunable Lyapunov}

The Lyapunov exponent is defined from four point function OTOC. To calculate the four-point function, one can solve the self-consistency equations for the four point functions in the real time to first obtain the asymptotic behavior in the chaos limit as shown in Ref.\cite{Kitaev1,Comments}, from which one can extract the Lyapunov exponent perfectly. We use $F_{\chi\chi}, F_{\psi\psi}$ and $F_{\chi\psi}$ to denote the four point functions of four $\chi$'s, of four $\psi$'s and of two $\chi$'s and two $\psi$'s. Let us consider the following OTOC in the real time, 
\begin{equation}
F_{\chi\chi}(t_1 , t_2)=\text{Tr}[y\chi_{i}(t_1)y\chi_{j}(0)y\chi_{i}(t_2)y\chi_{j}(0)],~~~~ y=\rho (\beta)^{\frac{1}{4}}.
\end{equation}
\begin{equation}
F_{\psi\psi}(t_1 , t_2)=\text{Tr}[y\psi_{i}(t_1)y\psi_{j}(0)y\psi_{i}(t_2)y\psi_{j}(0)],~~~~ y=\rho (\beta)^{\frac{1}{4}}.
\end{equation}
where fermions are separated by a quarter of the thermal circle, at the lowest order $\mathcal{O}(N^{0})$, the four point function is given by the disconnected part, and what we are interested in is the sub-leading part that describes chaotic behavior. Below, we use $F(t_1,t_2)$ to refer to the sub-leading part of the four point function without further indication. As in the original SYK model, they are determined by the ladder diagrams in the large-$N$ limit.

\begin{figure}[t]
	\centering
	\includegraphics[width=1\textwidth]{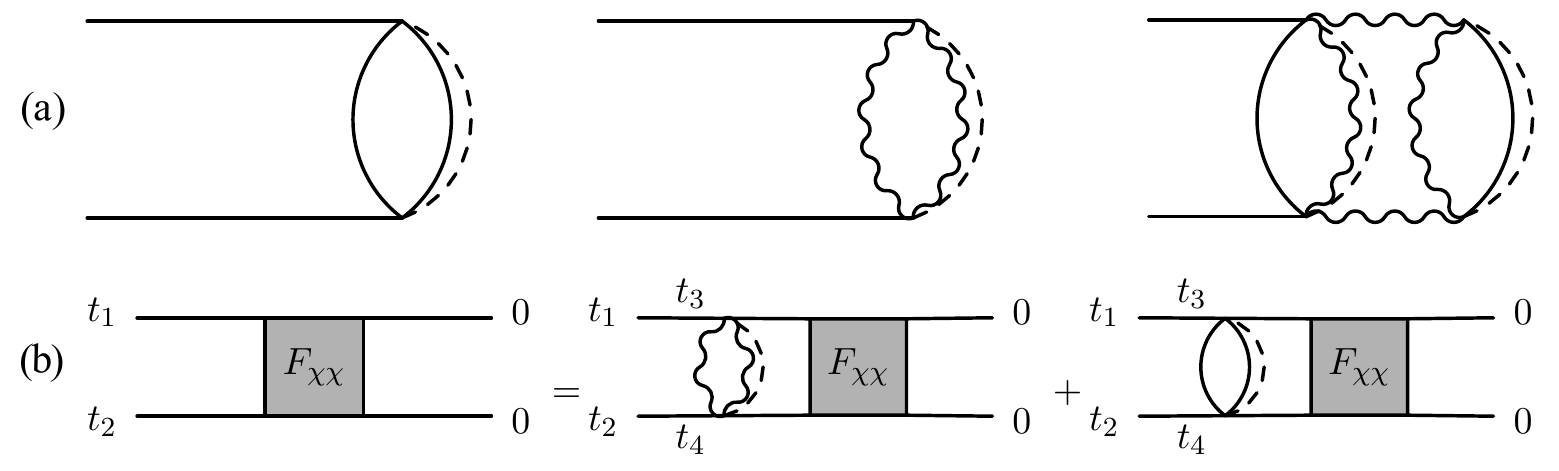}
	\caption{(a). Three contributions of $K_{R,\chi\chi}(t_1 ... t_4)$. The first two diagrams are of order $\frac{1}{N}$ while the third one is of order $\frac{1}{N^2}$. (b). Diagrammatic representation of the self consistency equation Eq. (\ref{eq10}).}\label{Feydia2}
\end{figure} 

We use self-consistency equations to determine the asymptotic behavior of $F_{\chi\chi}$, i.e., the function $F_{\chi\chi}$ is an eigenfunction of $K_{R,\chi\chi}$ with eigenvalue one:
\begin{equation}\label{eq10}
F_{\chi\chi}(t_1 , t_2)=\int dt_3 dt_4 K_{R,\chi\chi}(t_1 ... t_4)F_{\chi\chi}(t_3, t_4),
\end{equation}
where $K_{R,\chi\chi}(t_1 ... t_4)$ is the retarded kernel for evaluating the four point function. We shall first analyze the structure of the retarded kernel. One finds that the kernel consists of two parts at lowest order given by the first and second diagram in Fig. \ref{Feydia2} (a). Other possibilities, like the third diagram in Fig. \ref{Feydia2} (a), is in fact of the order $\frac{1}{N^2}$. At leading order for the connected part of four-point function, it can be neglected. From the diagrams, one can obtain the retarded kernel $K_{R,\chi\chi}(t_1 ... t_4)$ as
\begin{equation}\label{eq11}
K_{R,\chi\chi}(t_1 ... t_4)= 3J^2 G_{R,\chi}(t_{13})G_{R,\chi}(t_{24})G_{lr, \chi}(t_{34})^2 + u^2 G_{R,\chi}(t_{13})G_{R,\chi}(t_{24})G_{lr, \psi}(t_{34})^2,
\end{equation}
where $G_{R}(t)$ is the real time retarded correlator and $G_{lr}(t)\equiv iG(it+\frac{\beta}{2})$ is the Wightmann correlator. They are given by
\begin{equation}\label{eq12}
G_{R,\chi}(t)=\sqrt{2}a\theta (t) \left[\frac{\pi}{\beta \sinh \frac{\pi t}{\beta}}\right]^{\frac{1}{2}}, ~~~ G_{lr,\chi}(t)=a \left[\frac{\pi}{\beta \cosh \frac{\pi t}{\beta}}\right]^{\frac{1}{2}},
\end{equation}
\begin{equation}\label{eq13}
G_{R,\psi}(t)=\sqrt{2}b\theta (t) \left[\frac{\pi}{\beta \sinh \frac{\pi t}{\beta}}\right]^{\frac{1}{2}}, ~~~ G_{lr,\psi}(t)=b \left[\frac{\pi}{\beta \cosh \frac{\pi t}{\beta}}\right]^{\frac{1}{2}}.
\end{equation}

\begin{figure}[t]
\centering
\includegraphics[width=0.6\textwidth]{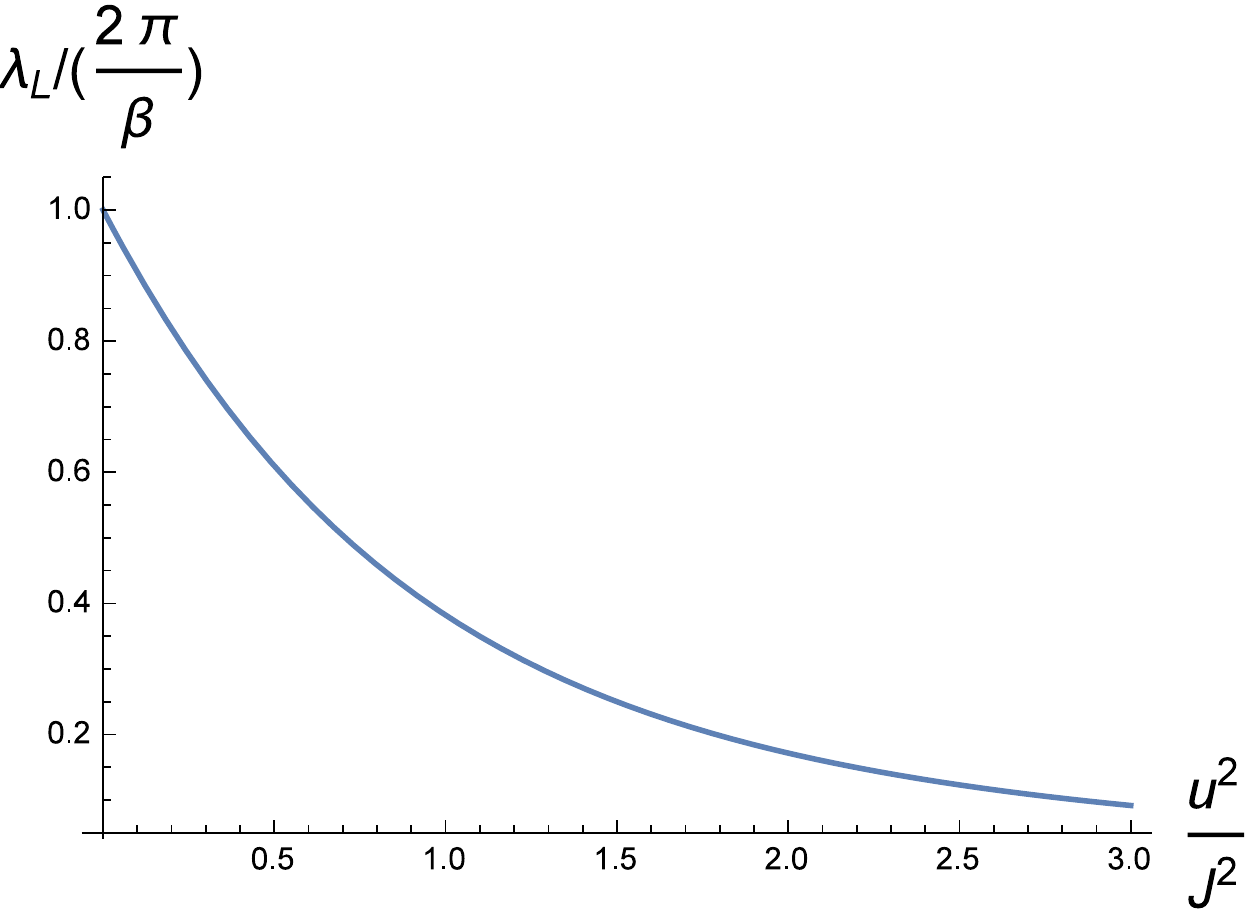}
\caption{The dependence of Lyapunov exponent of $F_{\chi\chi}$ on $u^2/J^2$.}\label{1}
\end{figure}

By taking an ansatz form of 
\begin{align}
F_{\chi\chi}(t_1,t_2)=\frac{e^{-h\frac{\pi}{\beta}(t_1 + t_2)}}{\left[\cosh \frac{\pi}{\beta}t_{12}\right]^{\frac{1}{2}-h}},
\end{align} 
and substituting Eq. (\ref{eq11}) - (\ref{eq13}) into Eq. (\ref{eq10}) while requiring that the eigenvalue of the kernel $k_{R}(h)=1$, one finds
\begin{equation}\label{eq14}
1=\frac{\Gamma (\frac{5}{2})\Gamma (\frac{1}{2}-h)}{\Gamma(\frac{3}{2})\Gamma(\frac{3}{2}-h)}\times 4\pi \left(J^2 a^4 + \frac{u^2 a^2 b^2}{3}\right).
\end{equation}
The value of the parenthesis in the above equation equals to $1/4\pi$ in the original SYK model, while here the value depends on $u$ explicitly and implicitly via coefficients $a$ and $b$. One can solve $h$ from Eq. (\ref{eq7}) and Eq. (\ref{eq14}), the result is
\begin{equation}
-h = 1 + \frac{1}{2}k^2 - \frac{\sqrt{k^4 + 4k^2}}{2},~~~ \text{with} ~~ k = \frac{u^2}{J^2}.
\end{equation}
One can see that Since $\lambda_{L}= -h\frac{2 \pi}{\beta}$, we find that
\begin{equation}
\lambda_L = \frac{2\pi}{\beta}\left(1 -\frac{\sqrt{k^4 + 4k^2}-k^2}{2}\right). \label{lambda}
\end{equation}
When $k=u^2/J^2=0$, the two systems decouple, and the Lyapunov exponent of $F_{\chi\chi}$ recovers $2\pi/\beta$, which is simply the maximumly chaotic value in the SYK model. However, as one increases the interaction between the two systems, i.e., $u^2/J^2$ increases, the Lyapunov exponent decreases. When $u^2/J^2\rightarrow \infty$, $\lambda_L\rightarrow 0$.  Fig. \ref{1} shows one of the central results of this work where we plot the dependence of $\lambda_L$ on $u^2/J^2$. For small $u^2/J^2$, the Lyapunov exponent decreases linearly with $u^2/J^2$:
\begin{equation}\label{eq18}
\lambda_L \approx \frac{2\pi}{\beta}\left(1-\frac{u^2}{J^2}\right),\,\,\,\, \frac{u^2}{J^2}\ll 1.
\end{equation}

The above calculation deals with the four point function in the real time and in the chaos limit directly. One can also first calculate the exact four point function in the imaginary time and then continue it to the real time, which will lead to the same result \cite{Wenbo}.

If we take the Lyapunov exponent $\lambda_L$ as a measurement of the chaos in one system, then this result tells us that we can tune chaotic behavior in one SYK system by changing the magnitude of its interaction with a much larger system. The underlying mechanism is based on the hierarchy of the scrambling times between these two systems. In the SYK model, the scrambling time $t_{s}$ is proportional to $\beta\log N$, where $N$ is the number of Majorana fermions in the system. The large difference in the size of the two subsystems results in a large difference in the scrambling time. At the time scale when the SYK$_\chi$ system enters the chaos region, the SYK$_\psi$ system has not scrambled at all. Thus, through the coupling between the two systems, the chaotic behavior of the SYK$_\chi$ system is weaken by the larger system SYK$_\psi$. 

\begin{figure}[t]
	\centering
	\includegraphics[width=0.8\textwidth]{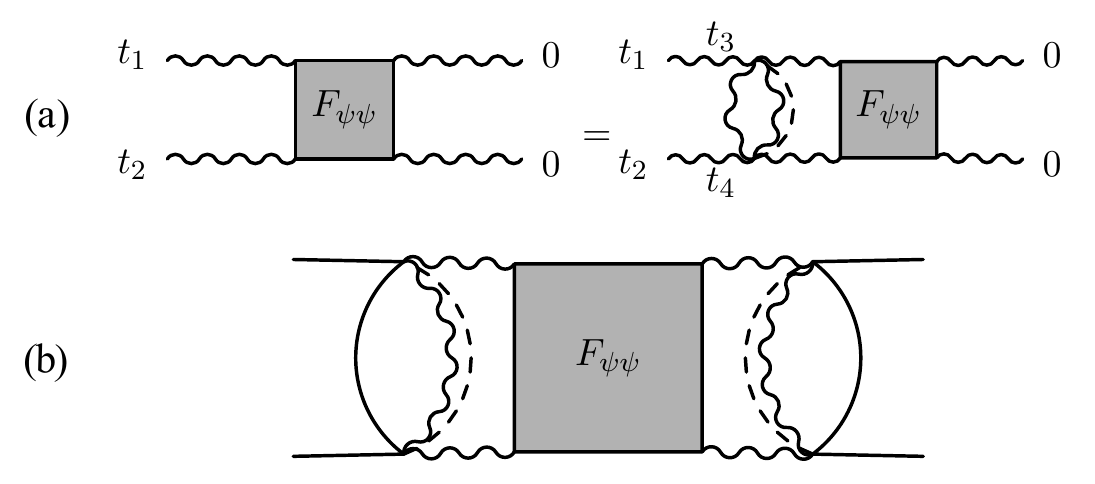}
	\caption{(a). Diagrammatic representation of the self consistency equation (\ref{eq18}). (b). At order $\mathcal{O}(\frac{1}{N^2})$, the OTOC $F_{\chi\chi}$ contains $F_{\psi\psi}$ as an inner propagator.}\label{Feydia3}
\end{figure}

Now let us turn to the OTOC $F_{\psi\psi}(t_1 , t_2)$ of four $\psi$ fermions. As illustrated in the last section, at the lowest order of $1/N$, the SYK$_{\psi}$ system should not be affected by the small system. We will find this is indeed the case here. One would find that the leading order of the connected part $F_{\psi\psi}(t_1 , t_2)$ is at order $\mathcal{O}(1/N^2)$, and the only contribution to the kernel at this order is shown in Fig. \ref{Feydia3} (a). From the self-consistency equation
\begin{equation}\label{eq18}
F_{\psi\psi}(t_1 , t_2)=\int dt_3 dt_4 K_{R,\psi\psi}(t_1 ... t_4)F_{\psi\psi}(t_3, t_4),
\end{equation}
and by taking the ansatz form 
\begin{equation}
F_{\psi\psi}(t_1,t_2)=\frac{e^{-h'\frac{\pi}{\beta}(t_1 + t_2)}}{\left[\cosh \frac{\pi}{\beta}t_{12}\right]^{\frac{1}{2}-h'}},
\end{equation}
with the eigenvalue $k_{R}(h')=1$, one finds
\begin{equation}
1=\frac{\Gamma (\frac{5}{2})\Gamma (\frac{1}{2}-h')}{\Gamma(\frac{3}{2})\Gamma(\frac{3}{2}-h')}\times 4\pi J^2 b^4 .
\end{equation}
Solving this one obtains $h'=-1$, which means that the Lyapunov exponent $\lambda_L$ is $2\pi/\beta$.
The Lyapunov exponent of $F_{\psi\psi}$ is the same as the one in the original SYK model, which matches our previous expectation.

If we want to understand the behavior of $F_{\chi\chi}$ at next order $\mathcal{O}(\frac{1}{N^2})$ precisely, we need to take into account the $\mathcal{O}(\frac{1}{N})$ correction to the correlators, and analyze the kernel structure as before. Some useful hint can be obtained by simply looking at the third diagram in Fig. \ref{Feydia1} (a), which is the $1/N^2$ order contribution to the kernel $K_{R,\chi\chi}$. It contains the four-point function $F_{\psi\psi}$ as a propagator inside (see Fig. \ref{Feydia3} (b)), and since the Lyapunov exponent of $F_{\psi\psi}$ reaches the maximal value $2\pi/\beta$, we anticipate that at order $1/N^2$, $F_{\chi\chi}(t_1 , t_2)$ will also display a Lyapunov exponent $2\pi/\beta$. This gives an exponential growth contribution as $\frac{1}{N^2}\exp(\frac{2\pi}{\beta}t)$ which reveals the maximally chaotic behavior at longer time when both systems are scrambled. Similarly, since the lowest order of the connected part of OTOC $F_{\psi\chi}$ is at $\mathcal{O}(\frac{1}{N^2})$ and it also contains $F_{\psi\psi}$ as an inner propagator, we anticipate that $F_{\psi\psi}$ will also grow as $\exp{\frac{2\pi}{\beta}t}$ at order $\frac{1}{N^2}$.

From this point of view, one can see different chaotic behavior at different time scale. As far as the small system is concerned, there are two scrambling time scales, with the first one at the shorter time and the second one at the longer time. The small system is not maximally chaotic at the first scrambling time, because of the non-chaotic environment at that time scale. And it is maximally chaotic at the second scrambling time at longer time scale, because of the maximally chaotic environment at that time scale.   

\subsection{The Effective Action \label{action}}

In this section we will discuss an alternative effective action derivation for the change of the Lyapunov exponent in a perturbative manner. This will be useful for the later discussion of the SYK chain coupled to the environment. 

The derivation of the effective action utilizes the original fermion path integral formalism and carries out the disorder average. By introducing bi-local fields $G_{\chi}(\tau_1,\tau_2), G_{\psi}(\tau_1,\tau_2)$ and Lagrange multiplier fields $\Sigma_{\chi}(\tau_1,\tau_2), \Sigma_{\psi}(\tau_1,\tau_2)$ that sets $G_{\chi}(\tau_1,\tau_2)=\frac{1}{N}\Sigma_{i}\chi_{i}(\tau_1)\chi_{i}(\tau_2)$ and $G_{\psi}(\tau_1,\tau_2)=\frac{1}{N^2}\Sigma_{i}\psi_{i}(\tau_1)\psi_{i}(\tau_2)$, one obtains the nonlocal action as
\begin{equation}
\begin{aligned}
S_{\textrm{eff}} = & N \left(-\frac{1}{2}\log \det (\partial_{\tau}-\Sigma_{\chi}) +\frac{1}{2}\int d\tau_1 d\tau_2 \left( \Sigma_{\chi}(\tau_1,\tau_2)G_{\chi}(\tau_1,\tau_2)-\frac{J^2}{4}G_{\chi}(\tau_1,\tau_2)^4  \right)\right)\\
& + N^2 \left(-\frac{1}{2}\log \det (\partial_{\tau}-\Sigma_{\psi}) +\frac{1}{2}\int d\tau_1 d\tau_2 \left( \Sigma_{\psi}(\tau_1,\tau_2)G_{\psi}(\tau_1,\tau_2)-\frac{J^2}{4}G_{\psi}(\tau_1,\tau_2)^4  \right)\right)\\
& -N\left(\frac{1}{2}\int d\tau_1 d\tau_2  \frac{u^2}{2} G_{\chi}(\tau_1,\tau_2)^2 G_{\psi}(\tau_1,\tau_2)^2 \right).
\end{aligned}
\end{equation}
The saddle point equations of the fields are the same as in Eq. (\ref{eq4}) and Eq. (\ref{eq5}). We expand the fields around their saddle point solution as
\begin{equation}
\begin{aligned}
&G_\chi (\tau)=G_{\chi,s}(\tau)+|G_{\chi,s}(\tau)|^{-1}g(\tau),\ \ \ \ \Sigma_\chi(\tau)  =\Sigma_{\chi,s}(\tau)+|G_{\chi,s}(\tau)|\sigma(\tau)\\
&G_\psi (\tau)=G_{\psi,s}(\tau)+|G_{\psi,s}(\tau)|^{-1}g'(\tau),\ \ \ \ \Sigma_\psi(\tau)  =\Sigma_{\psi,s}(\tau)+|G_{\psi,s}(\tau)|\sigma'(\tau),
\end{aligned}
\end{equation}
where $g(\tau)$, $g^\prime(\tau)$, $\sigma(\tau)$, $\sigma^\prime(\tau)$ denote the fluctuation of the fields.
By keeping up to the quadratic order, one obtains the effective action
\begin{equation}
\begin{aligned}
&S_{\text{eff}}=N\left(-\frac{1}{2}\log(1+G_{\chi,s}\circ G_{\chi,s}\sigma)+\int d\tau_1d\tau_2\frac{1}{2}g(\tau_1,\tau_2)\sigma(\tau_1,\tau_2)-\frac{3J^2}{4}g(\tau_1,\tau_2)g(\tau_1,\tau_2)\right)\notag\\
&+N^2\left(-\frac{1}{2}\log(1+G_{\psi,s}\circ G_{\psi,s}\sigma')+\int d\tau_1d\tau_2\frac{1}{2}g'(\tau_1,\tau_2)\sigma'(\tau_1,\tau_2)-\frac{3J^2}{4}g'(\tau_1,\tau_2)g'(\tau_1,\tau_2)\right)\notag\\
&-N\int d\tau_1d\tau_2 \left(u^2g'(\tau_1,\tau_2)g(\tau_1,\tau_2)+\frac{u^2}{4}g(\tau_1,\tau_2)g(\tau_1,\tau_2)+\frac{u^2}{4} g'(\tau_1,\tau_2)g'(\tau_1,\tau_2)\right).
\end{aligned}
\end{equation}
Integrating out $\sigma$ and $\sigma'$, and to the leading order of $\frac{u^2}{J^2}$, one obtains 
\begin{align}
&S_{\text{eff}}=\frac{3J^2N^2}{4}\int d\tau_1...d\tau_4g'(\tau_1,\tau_2)(K'^{-1}-\hat{1})g'(\tau_3,\tau_4)\notag\\
&+\frac{3J^2N}{4}\int d\tau_1...d\tau_4g(\tau_1,\tau_2)(K^{-1}-\hat{1})g(\tau_3,\tau_4)\notag\\&-N\int d\tau_1d\tau_2 \left( u^2g'(\tau_1,\tau_2)g(\tau_1,\tau_2)+\frac{u^2}{4}(g(\tau_1,\tau_2)g(\tau_1,\tau_2)+g'(\tau_1,\tau_2)g'(\tau_1,\tau_2))\right),
\end{align}
in which 
\begin{equation}
K'(\tau_1,\tau_2;\tau_3,\tau_4)=3J^2G_{\psi,s}(\tau_{12})G_{\psi,s}(\tau_{13})G_{\psi,s}(\tau_{42})G_{\psi,s}(\tau_{34}),
\end{equation}
and 
\begin{equation}
K=\frac{J^2}{J^2+u^2}K'.
\end{equation}
$K'$ does not have $J$ and $u$ dependences since the coefficient $b$ in $G_{\psi,s}(\tau)$ will cancel out the $J^2$ prefactor,  while $K$ has $J$ and $u$ dependences. 

The exact four point function can be derived by first diagonalizing the kernels $K$ and $K'$, with their eigenvalues denoted by $k(h,n)$ and $k'(h,n)$ and the eigenfunction corresponding to $(h,n)$ denoted by $\Psi_{h,n}$. For the SYK$_{\psi}$ system, at the leading order, one only needs to consider the first term in the effective action. The connected part of the imaginary time four-point function for SYK$_{\psi}$ system can be written as
\begin{align}
&\frac{1}{N^2}F_{\psi\psi}(\tau_1,\tau_2;\tau_3,\tau_4)    =\frac{1}{|G_{\psi,s}(\tau_{12})G_{\psi,s}(\tau_{34})|}\langle g'(\tau_1,\tau_2)g'(\tau_3,\tau_4)\rangle\\
& = \frac{2}{3N^2J^2}\frac{1}{|G_{\psi,s}(\tau_{12})G_{\psi,s}(\tau_{34})|}\left(K'^{-1}-\hat{1}\right)^{-1}\\
& = \frac{2}{3N^2J^2}\frac{1}{|G_{\psi,s}(\tau_{12})G_{\psi,s}(\tau_{34})|}\sum_{h,n}\Psi_{h,n}(\tau_1,\tau_2)\frac{k'(h,n)}{1-k'(h,n)}\Psi^{*}_{h,n}(\tau_3,\tau_4).
\end{align}
However, for $h=2$, one has $k'(h,n)=1$, which leads to the vanishing of effective action for $h=2$ modes as well as a divergence of the four-point function in above equation. The divergence comes from the reparametrization modes which are soft modes in the conformal limit. For them, one must consider the correction away from the conformal limit to obtain a meaningful result. On the other hand, these modes are the dominant part of the four-point function in the long time limit, and result in the exponential growth with the Lyapunov exponent $2\pi/\beta$ for the OTOC. This is the same story as the original SYK model, as well as the SYK$_{\psi}$ system. However, for the SYK$_{\chi}$ system, the interaction with the bath makes the difference. If we integrate out $g^\prime$ in the effective action, at the leading order, the effective action for $g$ is
\begin{equation}
S_{\textrm{eff},\chi}=\frac{3J^2 N}{4}\int d\tau_1 ...d\tau_4 g(\tau_1,\tau_2)\left(K^{-1}-\left(1+\frac{u^2}{3J^2}\right)\hat{1}\right)g(\tau_3,\tau_4).
\end{equation}
With this, one obtains the expression for the connected part of the imaginary time four-point function
\begin{align}
&\frac{1}{N}F_{\chi\chi}(\tau_1,\tau_2;\tau_3,\tau_4) \nonumber\\
& = \frac{2}{3NJ^2}\frac{1}{|G_{\chi,s}(\tau_{12})G_{\chi,s}(\tau_{34})|}\sum_{h,n}\Psi_{h,n}(\tau_1,\tau_2)\frac{k(h,n)}{1-(1+\frac{u^2}{3J^2})k(h,n)}\Psi^{*}_{h,n}(\tau_3,\tau_4)\\
 & = \frac{2}{3N(J^2+u^2)}\frac{1}{|G_{\chi,s}(\tau_{12})G_{\chi,s}(\tau_{34})|}\sum_{h,n}\Psi_{h,n}(\tau_1,\tau_2)\frac{k'(h,n)}{1-\frac{1+\frac{u^2}{3J^2}}{1+\frac{u^2}{J^2}}k'(h,n)}\Psi^{*}_{h,n}(\tau_3,\tau_4).
\end{align}
For $h=2$, $k'(h,n)=1$, but there is no divergence due to $\frac{u^2}{J^2}\neq 0$. Since the divergence is removed, the long time behavior of the four-point function should be a result of both $h=2$ and $h\neq 2$ modes. In principle, the asymptotic behavior of OTOC in the chaos limit contains contributions from the $h\neq 2$ modes. However, for $u^2/J^2\ll 1$, the reparametrization modes still compose the dominant part of the four-point function, thus we can first focus on the $h=2$ part, and then consider the correction from $h\neq 2$ part perturbatively. We will show below that this leads to the same result as in previous section for the perturbative regime $u^2/J^2\ll 1$.

Concentrating on the $h=2$ part, we need to write an explicit action for the reparametrization modes. For $u=0$, the SYK$_{\chi}$ and SYK$_{\psi}$ systems are decoupled, thus we should have two sets of reparametrization mode $\epsilon$ and $\epsilon'$, corresponding to the reparametrization of $G_{\chi}$ and $G_{\psi}$, respectively. For finite $u$, since $G_{\chi}$ and $G_{\psi}$ are coupled through Eq. (\ref{eq4}), they should be reparametrized together. However, since we are working in the $u^2/J^2\ll 1$ regime, we shall still keep two sets of reparametrization modes, and treat the $u^2$ term as the first order perturbation that couples the two sets of modes together. Explicitly, the two sets of reparametrization modes are introduced as:
\begin{equation}
\begin{aligned}
g'=\frac{1}{\beta J}\sum_{n}f_n\epsilon'_n,\ \ \ \ g=\frac{1}{\beta \sqrt{J^2+u^2}}\sum_{n} f_n\epsilon_n\approx\frac{1}{\beta J}\sum_{n}f_n\epsilon_n ,
\end{aligned}
\end{equation}
with functions $f_n$ defined in Ref.\cite{Comments} and $\int f_nf_{m}\propto|n|(n^2-1)\delta_{m,-n}$. The eigenvalues of $K'$ for the reparametrization modes are approximated by $1-|n|\frac{ \sqrt{2} \alpha_K }{2\pi \beta J}$, while the eigenvalues of $K$ for the reparametrization modes are approximated by $\frac{J^2}{J^2 + u^2}-|n|\frac{ \sqrt{2}\alpha_K }{2\pi \beta J}$, in which $\alpha_K$ is a numerical factor \cite{Comments}. The effective action is then given by:
\begin{equation}
\begin{aligned}
S= & \frac{1}{256\pi}\left(\sum_n\frac{N^2 \sqrt{2}\alpha_K }{\beta J}(\epsilon'_{-n}n^2(n^2-1)\epsilon'_n)
+\sum_n\frac{N \sqrt{2}\alpha_K }{\beta J}(\epsilon_{-n}n^2(n^2-1)\epsilon_n)\right.\notag\\
& \,\,\,\,\,\,\,\,\,\,\,\,\,\left. -\sum_n\frac{4N u^2 }{3J^2}(\epsilon'_{-n}|n|(n^2-1)\epsilon_n)+\sum_n\frac{2N u^2 }{3J^2}(\epsilon_{-n}|n|(n^2-1)\epsilon_n)\right),
\end{aligned}
\end{equation}
where we neglected a term proportional to $N$ for quadratic term of $\epsilon'$. To the lowest order for $\epsilon_{n}'$, the action is just Schwarzian derivative while for $\epsilon_{n}$ there is an additional term $\propto |n|(n^2-1)$.

The $h=2$ contribution to the leading order of the connected part four-point function of the SYK$_{\chi}$ system can be read from this effective action as :
\begin{equation}
\begin{aligned}
F_{\chi\chi,h=2}(\tau,\tau_{12},\tau_{34}) &=
\frac{8}{\pi}\sum_{n}\frac{e^{-in\tau}}{\frac{\sqrt{2}\alpha_K |n|}{\beta J} + \frac{2u^2}{3J^{2}}}\frac{1}{|n|(n^2 -1)}f_{n}(\tau_{12})f_{n}(\tau_{34})\\
& = \frac{16J}{\sqrt{2}\alpha_K}\sum_{n}\frac{e^{-in\tau}}{\frac{2\pi |n|}{\beta}+ \frac{4\pi  u^2}{3\sqrt{2}J\alpha_K} }\frac{1}{|n|(n^2 -1)}f_{n}(\tau_{12})f_{n}(\tau_{34}),
\end{aligned}
\end{equation}
where $\tau = (\tau_1+\tau_2 -\tau_3 -\tau_4)/2$. By analytical continuation with $\tau_{12} = \tau_{34}=\beta/2$ and $\tau=it$, we obtain the growing term in the long time as
\begin{equation}
\frac{F_{\chi\chi,h=2}(t)}{G(\frac{\beta}{2})^2}\simeq -\frac{4\pi J}{\sqrt{2}\alpha_K}\cdot \frac{e^{\frac{2\pi}{\beta}t}}{\frac{2\pi}{\beta}+\frac{4\pi  u^2}{3\sqrt{2}J\alpha_K}}.
\end{equation}
We see that the Lyapunov exponent is still $2\pi/\beta$, because we have not taken the $h\neq 2$ corrections into account yet. Here we have two small parameters $1/\beta J$ and $u^2/J^2$, which are both of order $\epsilon$. Suppose $F_{\chi\chi}$ can be written in the form as
\begin{align}
F_{\chi\chi}(t)  & \sim -\frac{1}{a_1 \epsilon }\exp\left[\frac{2\pi}{\beta}(1-\lambda_1 \epsilon + ... )t\right]\\
& = -\frac{e^{\frac{2\pi}{\beta}t}}{a_1 \epsilon} + \frac{2\pi}{\beta}\frac{\lambda_1}{a_1} t e^{\frac{2\pi}{\beta}t} +\mathcal{O}(\epsilon).
\end{align}
It is analyzed in Ref.\cite{Comments} that the $h\neq 2$ part gives a growing term $\frac{6\pi}{\beta}te^{\frac{2\pi}{\beta}t}$. By matching the above expansion, we can obtain
\begin{equation}
\lambda_1 \epsilon = \frac{3\sqrt{2}\alpha_K}{2\beta J}+\frac{u^2}{J^2},
\end{equation}
where the first term is the first order correction to the Lyapunov exponent away from the conformal limit, while the second term is due to the interaction with the thermal bath. In the conformal limit $\beta J\rightarrow \infty$, the Lyapunov exponent is given by
\begin{equation}
\lambda_L \approx \frac{2\pi}{\beta}\left(1-\frac{u^2}{J^2}\right),\,\,\,\frac{u^2}{J^2}\ll 1,
\end{equation}
which is consistent with the result Eq. (\ref{lambda}) and Eq. (\ref{eq18}) in the sec. \ref{Tunable Lyapunov}.

\section{(1+1)-d SYK Models Coupled to a Thermal Bath}

In this section we consider the (1+1)-dimensional generalization of the previous model. By generalizing to a one dimensional chain model, we can study richer physics including the spatial dependence of OTOC, the energy transport property and so on.  There are several proposals for generalizing the SYK model to higher dimensions \cite{Yingfei1,Yingfei2,high-D1,high-D2-con}, here we mainly follow Ref. \cite{Yingfei1,Yingfei2} and focus on the generalization to one dimensional chain model. Two different configurations of chain model will be discussed in this section. In both cases, we first prepare an one dimensional SYK chain, with $N$ Majorana fermions $\chi_{i,x}$ on each site, and prepare a SYK$_{\psi}$ system with $N^2$  Majorana fermions $\psi_i$. In the first configuration, we use the SYK$_{\psi}$ system as a \textit{global} bath, and couple every site of the SYK$_\chi$ chain to the SYK$_{\psi}$ system uniformly. In the second configuration, we use the SYK$_{\psi}$ system as a \textit{local} bath and attach the SYK$_{\psi}$ system to one end of the SYK$_\chi$ chain. The first model preserves the translational symmetry along the chain, while the second model does not.

\subsection{(1+1)-d SYK Chain Coupled to a Global Thermal Bath \label{chain1} }

\begin{figure}[t]
\centering
\includegraphics[width=0.9\textwidth]{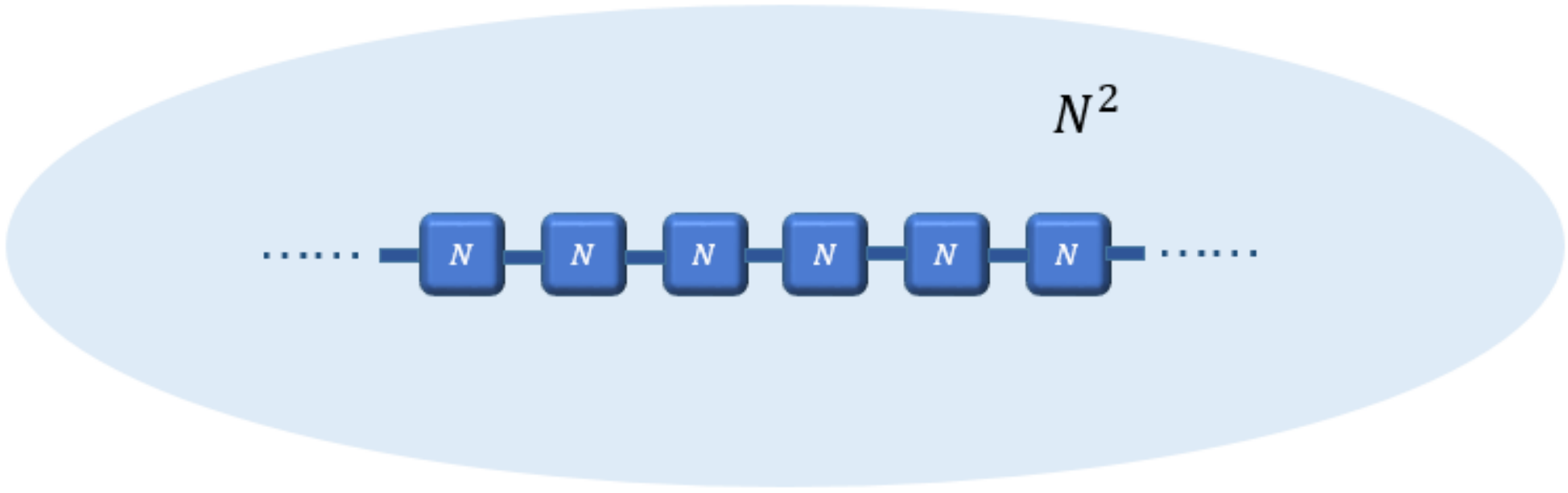}
\caption{A pictorial representation of the model discussed in sec. \ref{chain1}: An SYK chain with $N$ fermions each site in the bath of an $N^2$ fermions SYK model.}\label{9}
\end{figure}

In this subsection, we study the first configuration as illustrated in Fig. \ref{9}. To maintain the translational symmetry, here we choose the interaction strength with the $N^2$ fermions SYK$_{\psi}$ bath to be uniform. If the interaction strength is chosen different, this will be a new type of inhomogeneous SYK chain model, which can be analyzed using similar method as in Ref. \cite{Yingfei2}.

The Hamiltonian of the system is
\begin{equation}
H=\sum_{x=1}^{M}\left( H_{\chi_x} + H^{c}_{\chi_x,\chi_{x+1}} + H^{c}_{\chi_x,\psi} \right) + H_{\psi}\,\,\, ,
\end{equation}
where $H^{c}$ is defined as in Eq. (\ref{notation}). The random couplings in each term are $\{J_{jklm,x}\}$, $\{J^{'}_{jklm,x} \}$, $\{ u_{ijkl} \}$, $\{ \tilde{J}_{ijkl} \}$ in order,
with
\begin{equation}
\overline{J^{2}_{ijkl,x}}=\frac{3!J^{2}_{0}}{N^3},\,\,\,\overline{J^{'2}_{ijkl,x}}=\frac{J^{2}_{1}}{N^3},\,\,\, \overline{u_{ijkl,x}^{2}}=\frac{2\, u^2}{N^5},\,\,\, \overline{\tilde{J}_{ijkl}^{2}}=\frac{3!\tilde{J}^2}{N^6} .
\end{equation}
As we do not want the bath SYK$_{\psi}$ system to be affected by the existence of the chain at leading order of $\frac{1}{N}$, we demand the number of sites $M$ satisfy $\frac{M}{N}\rightarrow 0$ in the large-$N$ limit.

Using the replica method, after doing the disorder average and introducing the bi-local fields, one arrives at the effective action as (in below we omit the terms for the SYK$_{\psi}$ system):
\begin{equation}
\begin{aligned}
S_{\textrm{eff}}[G_x,\Sigma_x] & = \sum_{x=1}^{M}\left[ -\log{\textrm{Pf}(\partial_{\tau}-\Sigma_x)}+\frac{1}{2}\int_{0}^{\beta}d\tau_1 d\tau_2\left(\Sigma_x (\tau_1,\tau_2)G_{x}(\tau_1,\tau_2)-\frac{J_{0}^{2}}{4}G_{x}(\tau_1,\tau_2)^{4}     \right. \right.\\
& ~~~~~~~~~\left.\left. -\frac{J_{1}^{2}}{4}G_{x}(\tau_1,\tau_2)^2G_{x+1}(\tau_1,\tau_2)^2 - \frac{u^2}{2}G_{x}(\tau_1,\tau_2)^2G_{\psi}(\tau_1,\tau_2)^2  \right)\right]. 
\end{aligned}
\end{equation}
This effective action shows that the translational invariant saddle point solutions should satisfy:
\begin{equation}
G_{x}(\tau)=G^{s}(\tau),\,\, \Sigma_{x}(\tau)=\Sigma^{s}(\tau),
\end{equation}
\begin{equation}\label{eq28}
G^{s}(i\omega)^{-1}=-i\omega -\Sigma^{s}(i\omega),\,\, \Sigma^{s}(\tau)=\left(J_{0}^{2}+J_{1}^{2}+ u^2\frac{b^2}{a^2}\right)G^{s}(\tau)^{3},
\end{equation}
where $a$ and $b$ are the factors in front of the Green's functions $G^{s}(\tau)$ and $G_{\psi}(\tau)$ as in Eq. (\ref{eq6}). The value of $b^2/a^2$ can be tuned by changing the variance $\tilde{J}^2$ of the random couplings $\{\tilde{J}_{jklm}\}$. Here we set $b^2/a^2=1$ for the simplification of the equations by changing the value of $\tilde{J}^2$. We define
\begin{equation}
J=\sqrt{J_{0}^{2}+J_{1}^{2}+ u^2},
\end{equation}
then the Eq. (\ref{eq28}) becomes:
\begin{equation}\label{eq29}
G^{s}(i\omega)^{-1}=-i\omega -\Sigma^{s}(i\omega),\,\, \Sigma^{s}(\tau)=J^2G^{s}(\tau)^{3}.
\end{equation}
In the conformal limit $N\gg \beta J\gg 1$, one has the saddle point solution
\begin{equation}
\begin{aligned}
G^{s}(\tau) & =a \left[\frac{\pi}{\beta \sin \frac{\pi\tau}{\beta}}  \right]^{\frac{1}{2}},\,\,\, 0\leq \tau < \beta, \,\,\, \textrm{with}\,\,
a^4 J^2  = \frac{1}{4\pi}.
\end{aligned}
\end{equation}
Expanding the fields around the saddle point solutions as
\begin{equation}
G_{x}(\tau)=G^{s}(\tau)+|G^{s}(\tau)|^{-1}g_{x}(\tau),
\end{equation}
\begin{equation}
\Sigma_{x}(\tau)=\Sigma^{s}(\tau)+|G^{s}(\tau)|\sigma_{x}(\tau),
\end{equation}
one can expand the effective action to quadratic order as
\begin{equation}
\begin{aligned}
\delta S_{\textrm{eff}}[g,\sigma]=  & -\frac{1}{4}\int d^{4}\tau \sum_{x}\sigma_x(\tau_1 ,\tau_2)G^{s}(\tau_{13})\cdot |G^{s}(\tau_{34})|\cdot G^{s}(\tau_{42})\cdot|G^{s}(\tau_{21})|\sigma_{x}(\tau_3,\tau_{4})\\
& +\int d^{2}\tau \left(\sum_{x}\frac{1}{2}\sigma_{x}(\tau_1,\tau_2)g_{x}(\tau_{1},\tau_{2})-\frac{3J^2}{4}\sum_{x,y}g_{x}(\tau_1,\tau_2)S_{xy}g_{y}(\tau_1,\tau_2)\right).
\end{aligned}
\end{equation}
The spatial kernel $S_{xy}$ is defined as
\begin{equation}
S_{xy}=(1-\frac{2u^2}{3J^2})\delta_{x,y}+\frac{J^{2}_{1}}{3J^2}(\delta_{x,y\pm 1}-2\delta_{x,y}),
\end{equation}
with its Fourier transformation into momentum space being
\begin{equation}
s(p)=1-\frac{2u^2}{3J^2}+\frac{J^{2}_{1}}{3J^2}(\cos{p} -1).
\end{equation}

Defining $\tilde{K}$ as the symmetrized four-point function kernel of the SYK model
\begin{equation}
\tilde{K}(\tau_1,\tau_2;\tau_3,\tau_4)=3J^2 G^{s}(\tau_{13})\cdot |G^{s}(\tau_{34})|\cdot G^{s}(\tau_{42})\cdot|G^{s}(\tau_{21})|,
\end{equation}
and by integrating out the $\sigma_x$ fields, the effective action can be written as
\begin{equation}
\delta S_{\textrm{eff}}=\frac{3J^2}{4}\int d^{4}\tau \sum_{x,y}g_{x}(\tau_1,\tau_2)\left[ \tilde{K}^{-1}(\tau_1,\tau_2;\tau_3,\tau_4)\delta_{xy}-S_{xy}\delta(\tau_{13})\delta(\tau_{24}) \right]g_{y}(\tau_3,\tau_4).
\end{equation}
One can read the four point function with two fermions on the $x$ site and two fermions on the $y$ site from this action, which is
\begin{equation}
\begin{aligned}
\frac{1}{N}\mathcal{F}_{xy}(\tau_1,\tau_2;\tau_3,\tau_4) & =\frac{1}{|G^{s}(\tau_{12})G^{s}(\tau_{34})|}\langle g_{x}(\tau_1,\tau_2)g_{y}(\tau_3,\tau_4)\rangle\\
& = \frac{2}{3NJ^2}\frac{1}{|G^{s}(\tau_{12})G^{s}(\tau_{34})|}\left(\tilde{K}^{-1}-S\right)^{-1}.
\end{aligned}
\end{equation}
By Fourier transforming $(x-y)$ to the momentum space, one obtains the four point function in momentum space as
\begin{equation}
\frac{1}{N}\mathcal{F}_{p}(\tau_1,\tau_2;\tau_3,\tau_4)=\frac{2}{3NJ^2}\frac{1}{|G^{s}(\tau_{12})G^{s}(\tau_{34})|}\left[\tilde{K}^{-1}-s(p)\delta(\tau_{13})\delta(\tau_{24})\right]^{-1}.
\end{equation}
One can further get the behavior of the four point function from this expression by diagonalizing the kernel $\tilde{K}$ in the same way as in the original SYK model. The kernel has eigenvalues $k(h,n)$ and eigenfunctions $\Psi_{h,n}(\tau_1,\tau_2)$ which have been worked out in the Ref. \cite{Comments}. With these eigenvalues and eigenfunctions, an exact expression for the four-point function in the momentum space is given by
\begin{equation}
\frac{1}{N}\mathcal{F}_{p}(\tau_1,\tau_2;\tau_3,\tau_4)=\frac{2}{3NJ^2}\frac{1}{|G^{s}(\tau_{12})G^{s}(\tau_{34})|}\sum_{h,n}\Psi_{h,n}(\tau_1,\tau_2)\frac{k(h,n)}{1-s(p)k(h,n)}\Psi^{*}_{h,n}(\tau_3,\tau_4).
\end{equation}
The would-be divergence from $h=2$ is removed by $s(p)<1$, similar as in the section 2.3. Similarly, when the divergence has been removed, the long time behavior of the four point function comes from both the $h=2$ and $h\neq 2$ parts. Following the same procedure as in sec. \ref{action}, because for $p^2 \ll 1$ and $u^2/J^2\ll 1$, $s(p)\approx 1$, the four-point function in the long time and energy transport properties in the long wavelength limit are still mainly governed by the $h=2$ reparametrization modes. It is reasonable for us to first focus on the $h=2$ part, and then take into account of the corrections coming from $h\neq 2$ parts perturbatively when necessary. For the $h=2$ modes, we make reparametrizations to the saddle point solution
\begin{equation}
G^{s}(\tau_1,\tau_2)
\rightarrow G^{f}_{x}(\tau_1,\tau_2)=\left(f^{'}_{x}(\tau_1)f^{'}_{x}(\tau_2)\right)^{\frac{1}{4}}G^{s}(f_{x}(\tau_1),f_{x}(\tau_2)),
\end{equation}
with small deformations $f_{x}(\tau)=\tau+\epsilon_{x}(\tau)$. Using the same method as in sec. \ref{action}, we derive the effective action of the chain model in terms of the reparametrization modes:
\begin{equation}
\begin{aligned}
S=\frac{1}{256\pi}\sum_{n,p}\epsilon_{n,p}\left(\frac{\sqrt{2}\alpha_K}{\beta J}n^2 (n^2 -1)+\left(\frac{J^{2}_{1}}{3J^2}p^2 + \frac{2u^2}{3J^{2}}\right)|n|(n^2 -1)\right)\epsilon_{-n,-p}
\end{aligned}
\end{equation}
where $\epsilon_{n,p}$'s are the reparametrization modes by Fourier transforming $\epsilon_{x}(\tau)$:
\begin{equation}
\epsilon_{n,p}=\frac{1}{\sqrt{M}}\sum_{x=1}^{M}\int_{0}^{\beta}d\tau e^{i(\frac{2\pi n}{\beta}\tau - xp)}\epsilon_{x}(\tau).
\end{equation}
Thus the $h=2$ contribution to the four-point function can be derived as ($\tau=(\tau_1+\tau_2 - \tau_3 -\tau_4)/2$):
\begin{equation}\label{eq45}
\begin{aligned}
\frac{\mathcal{F}_{p,h=2}(\tau, \tau_{12},\tau_{34})}{G^{s}(\tau_{12})G^{s}(\tau_{34})} & = \frac{8}{\pi}\sum_{n}\frac{e^{-in\tau}}{\frac{\sqrt{2}\alpha_K |n|}{\beta J}+\frac{J^{2}_{1}}{3J^2}p^2 + \frac{2u^2}{3J^{2}}}\frac{1}{|n|(n^2 -1)}f_{n}(\tau_{12})f_{n}(\tau_{34})\\
& = \frac{16J}{\sqrt{2}\alpha_K}\sum_{n}\frac{e^{-in\tau}}{\frac{2\pi |n|}{\beta}+ \frac{1}{\tau_u} +D p^2}\frac{1}{|n|(n^2 -1)}f_{n}(\tau_{12})f_{n}(\tau_{34})
\end{aligned}
\end{equation}
in which
\begin{equation}
\frac{1}{\tau_u}\equiv\frac{4\pi  u^2}{3\sqrt{2}J\alpha_K},\,\,\, D\equiv\frac{2\pi  J^{2}_{1}}{3\sqrt{2}J\alpha_K}.
\end{equation}
$D$ is the diffusion constant of the system. To see this, one needs to consider the energy transport property of the system. The diffusive dynamics in space-time is governed by the reparameterization fields, which are the same as the ones that govern the long time behavior of fermion four-point function. The retarded energy-density correlation function in the $(p,\omega)$ space can be derived by utilizing the result of fermion four-point function in the OPE limit $|\tau_2 - \tau_1|,|\tau_4 - \tau_3|\ll \tau$. By the same method as in Ref. \cite{Comments}\cite{Yingfei1}, we get the expression for the retarded energy-density correlation function $C^{R}_{00}(p,\omega)$:
\begin{equation}
C^{R}_{00}(p,\omega)\simeq -\frac{Nc_v}{\beta} \frac{\frac{1}{\tau_u}+Dp^2}{-i\omega +\frac{1}{\tau_u} +Dp^2}.
\end{equation}
The denominator $(-i\omega +\frac{1}{\tau_u} +Dp^2)$ comes from a diffusion-type equation in real space
\begin{equation}
\partial_t \phi(x,t) = D\partial_{x}^{2}\phi(x,t) - \frac{1}{\tau_u} \phi(x,t),
\end{equation}
where $\phi(x,t)$ is the energy density function. From this equation, one finds the physical meaning of $D$ and $\tau_u$, where $D$ is the diffusion constant, while the $\frac{1}{\tau_u} \phi$ term will lead to an exponential decay of the energy density function, with $\tau_{u}\sim \frac{J}{u^2}$ being the characteristic time scale for the decay. Physically, this characterizes the process of energy leaking into the bath SYK$_{\psi}$ system through interaction, which is similar to the process that the energy flow crossing the black hole horizon being absorbed into the black hole, and can no longer be extracted by an exterior observer.

After the discussion of the energy transport property, we now turn to discuss the behavior of the OTOC in the chaos limit. The OTOC as a function of both space and time contains two majorana fermions from position $x$ and two from position $0$:
\begin{equation}
F(x , t)=\frac{1}{N^2}\sum_{i,j}^{N}\text{Tr}[y\chi_{i,x}(t)y\chi_{j,0}(0)y\chi_{i,x}(t)y\chi_{j,0}(0)],~~~~ y=e^{-\beta\hat{H}/4},
\end{equation}
from which we can extract both the Lyapunov exponent and the butterfly velocity. At the leading order $\mathcal{O}(N^0)$, the four-point function is given by a disconnected part $-G(\frac{\beta}{2})^2$, and the $\mathcal{O}(\frac{1}{N})$ part of the function can be derived by first doing analytical continuation of the Eq. (\ref{eq45}) with $\tau_{12} = \tau_{34}=\beta/2$ and $\tau=it$, and then Fourier transforming back to real space. By analytical continuation, we obtain the growing term in the long time as
\begin{equation}
\frac{\mathcal{F}_{p,h=2}(t)}{G(\frac{\beta}{2})^2}\simeq -\frac{4\pi J}{\sqrt{2}\alpha_K}\cdot \frac{e^{\frac{2\pi}{\beta}t}}{\frac{2\pi}{\beta}+\frac{1}{\tau_u}+Dp^2}.
\end{equation}
We see that it is exponential growing in the long time. In the same spirit as in sec. \ref{action}, we need to consider the $h\neq 2$ contributions to get the modified Lyapunov exponent. By same method, after taking into account the first order correction of $p^2, u^2/J^2$, as well as $1/(\beta J)$, one finds
\begin{equation}\label{eq3.28}
\frac{\mathcal{F}_{p}(t)}{G(\frac{\beta}{2})^2}\simeq -\frac{4\pi J}{\sqrt{2}\alpha_K}\cdot \frac{1}{\frac{2\pi}{\beta}+\frac{1}{\tau_u}+Dp^2}\exp\left[\frac{2\pi}{\beta}\left(1-\delta(p) \right)t\right].
\end{equation}
where 
\begin{equation}
\delta(p)=\frac{u^2}{J^2}+\frac{J^{2}_{1}}{2J^2}p^2+\frac{3\sqrt{2}\alpha_K}{2\beta J}.
\end{equation}
 $F(x,t)$ is given by Fourier transforming Eq. (\ref{eq3.28})  back into real space
\begin{equation}
\frac{F(x,t)}{-G(\frac{\beta}{2})^2}=1-\frac{1}{N}\frac{4\pi J}{\sqrt{2}\alpha_K}\int_{-\infty}^{\infty}\frac{dp}{2\pi}\frac{e^{ipx}}{\frac{2\pi}{\beta}+\frac{1}{\tau_u}+Dp^2}\exp\left[\frac{2\pi}{\beta}\left(1-\delta(p) \right)t\right]+\mathcal{O}(\frac{1}{N^2}).
\end{equation}
The residue at the pole $p^{*}=i\sqrt{\frac{\frac{2\pi}{\beta}+\frac{1}{\tau_u}}{D}}$ dominates the integral for large $x$. One obtains
\begin{equation}
\frac{F(x,t)}{-G(\frac{\beta}{2})^2}\simeq 1-\frac{1}{N}\frac{\sqrt{2}\pi J}{\alpha_K\sqrt{D(\frac{2\pi}{\beta}+\frac{1}{\tau_u})}}\exp\left[\frac{2\pi}{\beta}(t-\frac{x}{v_B})\right].
\end{equation}
We find that for large $x$, the first order correction to the Lyapunov exponent exactly cancels each other, leaving a Lyapunov exponent $2\pi/\beta$ unchanged as in Ref. \cite{Yingfei1}. However, this will not hold for small $x$ and the Lyapunov exponent will be tuned. The butterfly velocity is given by 
\begin{equation}\label{eq56}
v_B = \frac{\frac{2\pi}{\beta}}{\sqrt{\frac{2\pi}{\beta}+\frac{1}{\tau_u}}}\sqrt{D}.
\end{equation}
Defining $\tau_L=\frac{\beta}{2\pi}$, we have
\begin{equation}\label{eq57}
v^{2}_{B}=\frac{1}{\tau_L}\frac{\tau_u}{\tau_u + \tau_L} D. 
\end{equation}
This relationship is very interesting. On one hand, it still satisfies 
\begin{equation}
D\gtrsim v^{2}_{B} \tau_L,
\end{equation}
which is proposed to be a bound for the butterfly velocity \cite{vb bound,vb energy1,vb energy2,thermal transport,holographic review,Sachdev new,Blake}. It is found to held in a SYK chain \cite{Yingfei1} and holographic theories \cite{vbbh}.

\begin{figure}[t]
	\centering
	\includegraphics[width=0.7\textwidth]{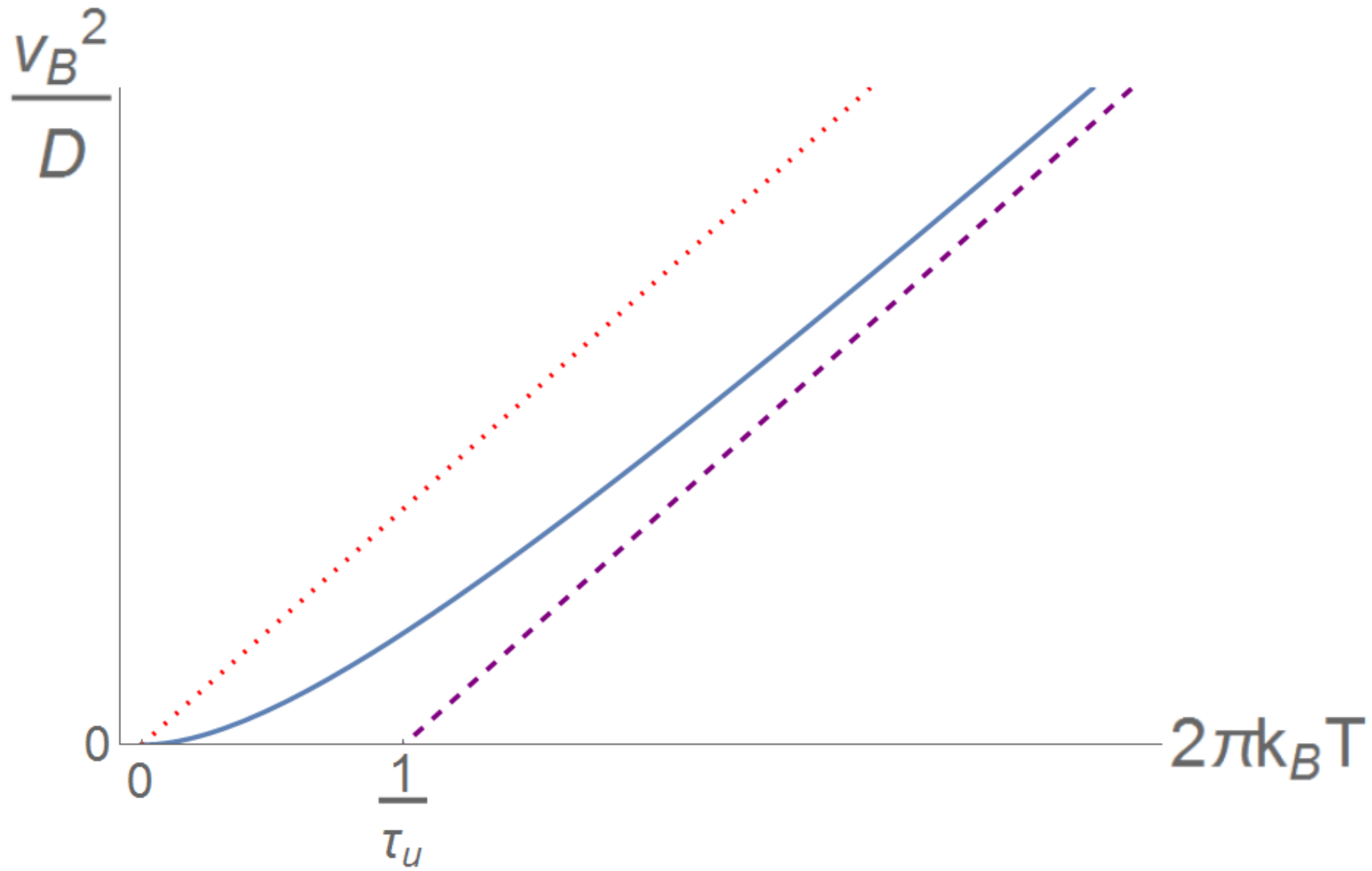}
	\caption{A sketch of $\frac{v_{B}^{2}}{D}-T$ in this model. The dotted line represents the bound proposed in the Ref. \cite{vb bound,vb energy1,vb energy2,thermal transport,holographic review,Sachdev new,Blake}. The solid curve shows the result of Eq. \eqref{eq57} and the dashed line gives its asymptotic behavior. The crossover occurs near $T\sim1/\tau_u$.}
	\label{vb}
\end{figure}

On the other hand, the scaling law of $v_{B}^{2}/D$ shows a crossover behavior as temperature increases, as shown in Fig. \ref{vb}. As we mentioned, $\tau_u$ is the time scale which governs the decay of the energy density function, while it can also define a temperature scale that is proportional to $u^2/J$. When the temperature of the system is much higher than this temperature scale, i.e., $k_B T \gg u^2/J$, Eq. (\ref{eq57}) tells 
\begin{equation}
\frac{v^{2}_{B}}{D}\propto T,
\end{equation}
which is the same temperature scaling law as in Ref. \cite{Yingfei1,Yingfei2}. However, when the temperature of the system is far below that temperature scale, i.e., $k_B T \ll u^2/J$, Eq. (\ref{eq57}) tells
\begin{equation}
\frac{v^{2}_{B}}{D}\propto T^2.
\end{equation}
This scaling law is similar as that in the Landau's Fermi liquid theory, in which $v_B$ is replaced by Fermi velocity $v_F$, which is a constant in low temperature, while $D$ scales as $\frac{1}{T^2}$. However, the difference is that here $D$ is a constant that does not depend on the temperature, while $v_B$ scales as $T$.

To conclude this subsection, we find that, by coupling the one dimensional SYK chain to a larger SYK bath, the butterfly velocity can not only acquire different values, but also acquire different temperature dependence. From Eq. (\ref{eq56}), it can be seen that when we increase the interaction strength $u$ with the bath, the butterfly velocity decreases, which physically means that the information leaks into the bath. Meanwhile, the dependence of the butterfly velocity on temperature transits from $v_B \propto \sqrt{T}$ to $v_B \propto T$.

\subsection{(1+1)-d SYK Chain Coupled to a Local Thermal Bath \label{chain2}}

\begin{figure}[t]
\centering
\includegraphics[width=0.9\textwidth]{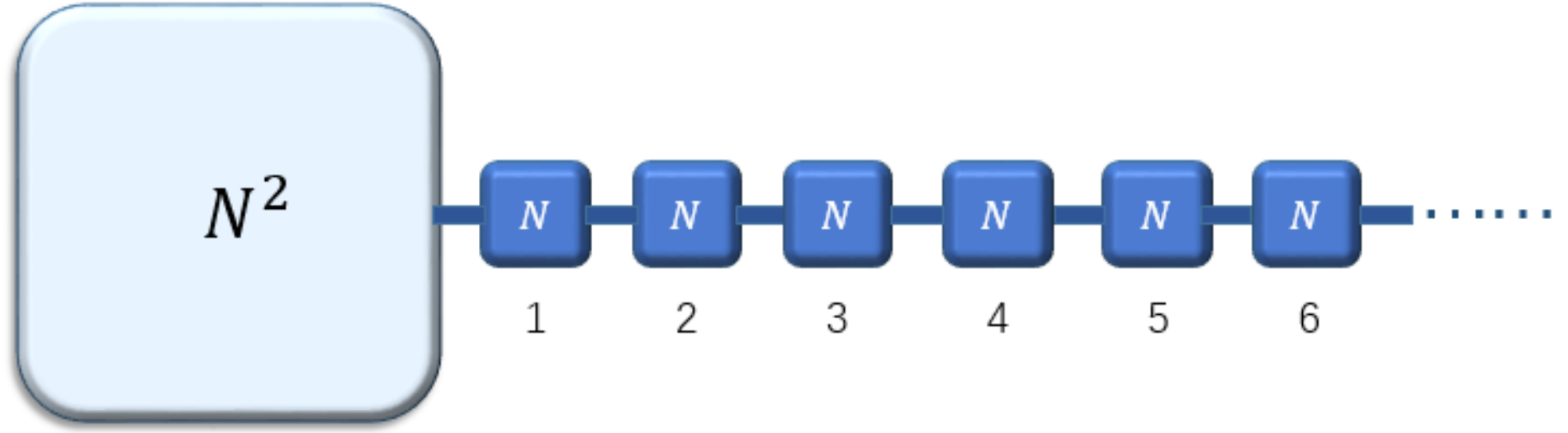}
\caption{A pictorial representation of the model discussed in sec. \ref{chain2}: An SYK chain with $N$ fermions each site in contact with a SYK system with $N^2$ fermions on one side.}\label{config2}
\end{figure}

In this subsection, we study another configuration, in which we couple the SYK$_{\psi}$ system to one end of the SYK chain as shown in Fig. \ref{config2}. Different from the model in sec. \ref{chain1}, this model is no longer translational invariant. In this configuration, our focus will be the spatial dependence of both the Lyapunov exponent and the butterfly velocity. We call the site which is nearest to SYK$_{\psi}$ as the site $1$, the second nearest as the site $2$, and so on.

The Hamiltonian of the system can be written as
\begin{equation}
H=\sum_{x=1}^{M}\left( H_{\chi_x} + H^{c}_{\chi_x,\chi_{x+1}}\right)+H^{c}_{\chi_1,\psi} + H_{\psi}\,\,\, .
\end{equation}
The random couplings in each term are $\{J_{ijkl,x}\},\{J^{'}_{ijkl,x}\},\{J^{'}_{ijkl,0}\},\{\tilde{J}_{ijkl}\}$, with
\begin{equation}\label{eq62}
\overline{J^{2}_{ijkl,x}}=\frac{3!J^{2}}{N^3},\,\,\,\overline{J^{'2}_{ijkl,x}}=\frac{2 \,u^{2}}{N^3},\,\,\, \overline{J^{'2}_{ijkl,0}}=\frac{2\, u^2}{N^5},\,\,\, \overline{\tilde{J}_{ijkl}^{2}}=\frac{3!J^2}{N^6} .
\end{equation}
Note that in Eq. (\ref{eq62}), the meanings of the symbols $J$ and $u$ are different from their meanings in the last subsection. For the physical picture to be clear, we only kept two parameters $u$ and $J$, although we could have four different parameters in the Eq. (\ref{eq62}). The Schwinger-Dyson equations are
\begin{equation}
G_{\psi}(i\omega)^{-1}=-i\omega - \Sigma_{\psi} (i\omega), \,\,\, G_{x}(i\omega)^{-1}=-i\omega - \Sigma_{x} (i\omega),
\end{equation}
with
\begin{equation}
\left\{ 
\begin{aligned}
\Sigma_{\psi}(\tau) & = J^2 G_{\psi}(\tau)^3 , \\ 
\Sigma_{1}(\tau) & = J^2 G_{1}(\tau)^3 + u^2 G_{1}(\tau)G_{\psi}(\tau)^2 + u^2 G_{1}(\tau)G_{2}(\tau)^2 , \\
\Sigma_{x}(\tau) & = J^2 G_{x}(\tau)^3 + u^2 G_{x}(\tau)G_{x-1}(\tau)^2 + u^2 G_{x}(\tau)G_{x+1}(\tau)^2 ,\,\, x\geq 2.
\end{aligned}
\right.
\end{equation}
Suppose the solutions in the conformal limit still obey the ansatz
\begin{equation}
G_{\psi}(\tau)= b\frac{\text{sgn}(\tau)}{|\tau|^{\frac{1}{2}}},~~~~ G_{x}(\tau)= a_x\frac{\text{sgn}(\tau)}{|\tau|^{\frac{1}{2}}},
\end{equation}
then one can get a set of coupled equations for the coefficients $b$ and $\{a_x\}$ as:
\begin{equation}
\left\{ 
\begin{aligned}
& J^2 b^4  =\frac{1}{4\pi} , \\ 
& J^2 a^{4}_{1}  + u^2 (a_{1}^{2} b^2 + a_{1}^{2}a_{2}^{2}) =\frac{1}{4\pi} , \\
& J^2 a^{4}_{x}  + u^2 (a_{x}^{2} a_{x-1}^{2} + a_{x}^{2}a_{x+1}^{2}) =\frac{1}{4\pi},\,\, x\geq 2.
\end{aligned}
\right.
\end{equation}
With the asymptotic boundary condition $a_x\rightarrow \textrm{const}$ when $x\rightarrow \infty$, one can solve the coupled equations numerically to get the coefficients $b$ and $\{a_x\}$.

Our goal is to calculate the OTOC functions in the chaos limit, and extract the Lyapunov exponents and butterfly velocities from them. In this model, the OTOCs $F_{1}(t_1,t_2)$, $F_{2}(t_1,t_2),...$ \footnote{$F_{i}(t_1,t_2)$ is the $\frac{1}{N}$ piece of the OTOC with four fermions from site $i$, as defined before.} are coupled to each other. For $F_{1}(t_1,t_2)$, the self-consistency equation involves $F_{2}(t_1,t_2)$:
\begin{equation}
F_{1}(t_1,t_2)=\int dt_3 dt_4 K_{11}(t_1 ... t_4)F_{1}(t_3, t_4)+K_{12}(t_1 ... t_4)F_{2}(t_3, t_4),
\end{equation}
where 
\begin{align}
K_{11}(t_1...t_4)& = 3J^2 G_{1,R}(t_{13})G_{1,R}(t_{24})G_{1,lr}(t_{34})^2\nonumber\\
&+u^2 G_{1,R}(t_{13})G_{1,R}(t_{24}) \left(G_{2,lr}(t_{34})^2+G_{\psi,lr}(t_{34})^2\right),
\end{align}
and
\begin{equation}
K_{12}(t_1...t_4)=2u^2 G_{1,R}(t_{13})G_{1,R}(t_{24})G_{1,lr}(t_{34})G_{2,lr}(t_{34}).
\end{equation}
For $F_{x}(t_1,t_2)$ with $x\geq 2$, the self-consistency equation involve $F_{x-1}(t_1,t_2)$ and $F_{x+1}(t_1,t_2)$:
\begin{align}
&F_{x}(t_1,t_2)=\nonumber\\
&\int dt_3 dt_4 K_{xx}(t_1 ... t_4)F_{1}(t_3, t_4)+K_{x,x-1}(t_1 ... t_4)F_{x-1}(t_3, t_4)+K_{x,x+1}(t_1 ... t_4)F_{x+1}(t_3, t_4),
\end{align}
with
\begin{align}
K_{xx}(t_1...t_4)&=3J^2 G_{x,R}(t_{13})G_{x,R}(t_{24})G_{x,lr}(t_{34})^2\nonumber\\
&+u^2 G_{x,R}(t_{13})G_{x,R}(t_{24}) \left(G_{x-1,lr}(t_{34})^2+G_{x+1,lr}(t_{34})^2\right),
\end{align}
and
\begin{equation}
K_{x,x\pm 1}(t_1...t_4)=2u^2 G_{x,R}(t_{13})G_{x,R}(t_{24})G_{x,lr}(t_{34})G_{x\pm 1,lr}(t_{34}).
\end{equation}
These equations can be casted into a matrix form with the kernel matrix $\mathcal{K}=\{K_{ij}\}$, and by solving the matrix integral equation, one would get different modes with different Lyapunov exponents, similar as different momentum modes in sec. \ref{chain1}. The matrix integral equation can be solved by the ansatz solution as
\begin{equation}
\vec{\mathcal{F}_h}\frac{e^{-h\frac{\pi}{\beta}(t_1 + t_2)}}{\left[\cosh \frac{\pi}{\beta}t_{12}\right]^{\frac{1}{2}-h}}=\int dt_3 dt_4 \mathcal{K}(t_1 ... t_4)\vec{\mathcal{F}_h}\frac{e^{-h\frac{\pi}{\beta}(t_3 + t_4)}}{\left[\cosh \frac{\pi}{\beta}t_{34}\right]^{\frac{1}{2}-h}}.
\end{equation}
From this one can solve a series of eigen-modes, each with different $h$, which corresponds to different Lyapunov exponents. A technical detail is that the kernel matrix $\mathcal{K}=\{K_{ij}\}$ is non-Hermitian, thus the eigen-modes $\{\vec{\mathcal{F}_h}\}$ are not orthogonal with each other. To keep the eigen-modes orthogonal with each other, one needs to use a symmetrized kernel matrix $\tilde{\mathcal{K}}=\{\tilde{K}_{ij}\}$. After getting a series of different modes, the $\frac{1}{N}$ part of the OTOC $\langle \chi_{i,x} (t)\chi_{j,x}(0)\chi_{i,x} (t)\chi_{j,x}(0) \rangle_\beta$ with four fermions on the $x$ site can be written as a summation of different modes:
\begin{equation}\label{eq81}
F_{x}(t)\propto\sum_{h}\tilde{\mathcal{F}}^2_{hx}\frac{1}{\frac{2\pi}{\beta}+\alpha (h+1)}\exp\left[-h\frac{2\pi}{\beta}t\right],
\end{equation}
where $\alpha = \frac{4\pi J}{3\sqrt{2}\alpha_K}$. The coefficients $\{\tilde{\mathcal{F}}_{hx}\}$ are the mode expansion coefficients derived in numerics by using the symmetrized kernel matrix. The weight of different modes in summation should be determined by the effective action, while one can also compare with the results in the last subsection, and see it should be approximated by $\frac{1}{\frac{2\pi}{\beta}+\alpha (h+1)}$. For the OTOC $\langle \chi_{i,x} (t)\chi_{j,y}(0)\chi_{i,x} (t)\chi_{j,y}(0) \rangle_\beta$ with two fermions on the $x$ site and two fermions on the $y$ site, it can be derived as
\begin{equation}
F_{xy}(t)\propto\sum_{h}\tilde{\mathcal{F}}_{hx}\tilde{\mathcal{F}}_{hy}\frac{1}{\frac{2\pi}{\beta}+\alpha (h+1)}\exp\left[-h\frac{2\pi}{\beta}t\right].
\end{equation}

\begin{figure}[t]
\centering
\includegraphics[width=1\textwidth]{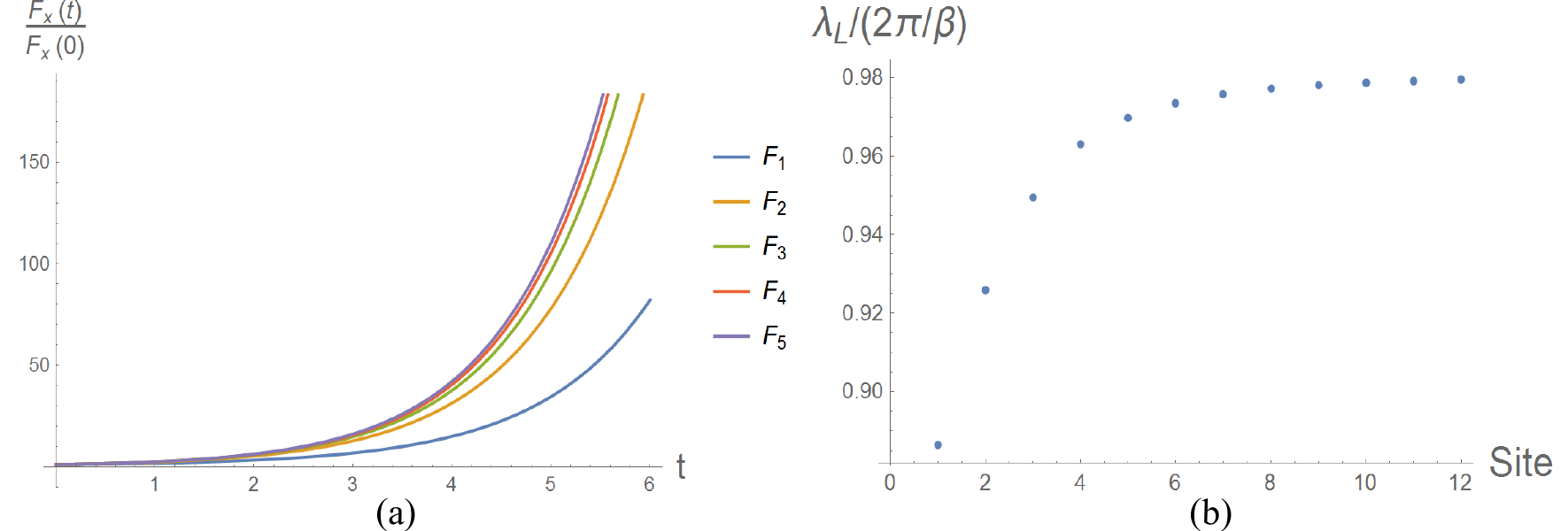}
\caption{(a) Different behavior of $F_{1}(t) \sim F_{5}(t)$. (b) Fitted Lyapunov exponent from the behavior of $F_{x}(t)$ after the dissipation time $\beta/2\pi$. $\beta =2\pi$, $J=u=100$ are chosen.}\label{spatial}
\end{figure}

We first consider the OTOC with four fermions from the same site. We give the plot of the behavior of $F_{x}(t)$ for the first five sites away from the SYK$_{\psi}$ system in the Fig. \ref{spatial} (a). In numerics, we set $\beta = 2\pi$, and $J=u=100$. We can see that, the closer to the SYK$_{\psi}$ system, the slower the growth of $F_{x}(t)$. To quantify the growth rate of $F_{x}(t)$, we fit $F_{x}(t)$ after the dissipation time $t_{d}\sim\frac{\beta}{2\pi}$ with function $e^{\lambda_L t}$. We get the local Lyapunov exponent $\lambda_L$ for each site, which characterizes the growth rate of the OTOC. The result is shown in Fig. \ref{spatial} (b). We see that the Lyapunov exponents acquire a non-trivial spatial dependence. In our (0+1)-d model, the interaction with a much larger cluster of SYK system results in the decrease of Lyapunov exponent of the small system. In this (1+1)-d model, this influence depends on the distance between the large system and each site on the chain.

\begin{figure}[t]
\centering
\includegraphics[width=0.5\textwidth]{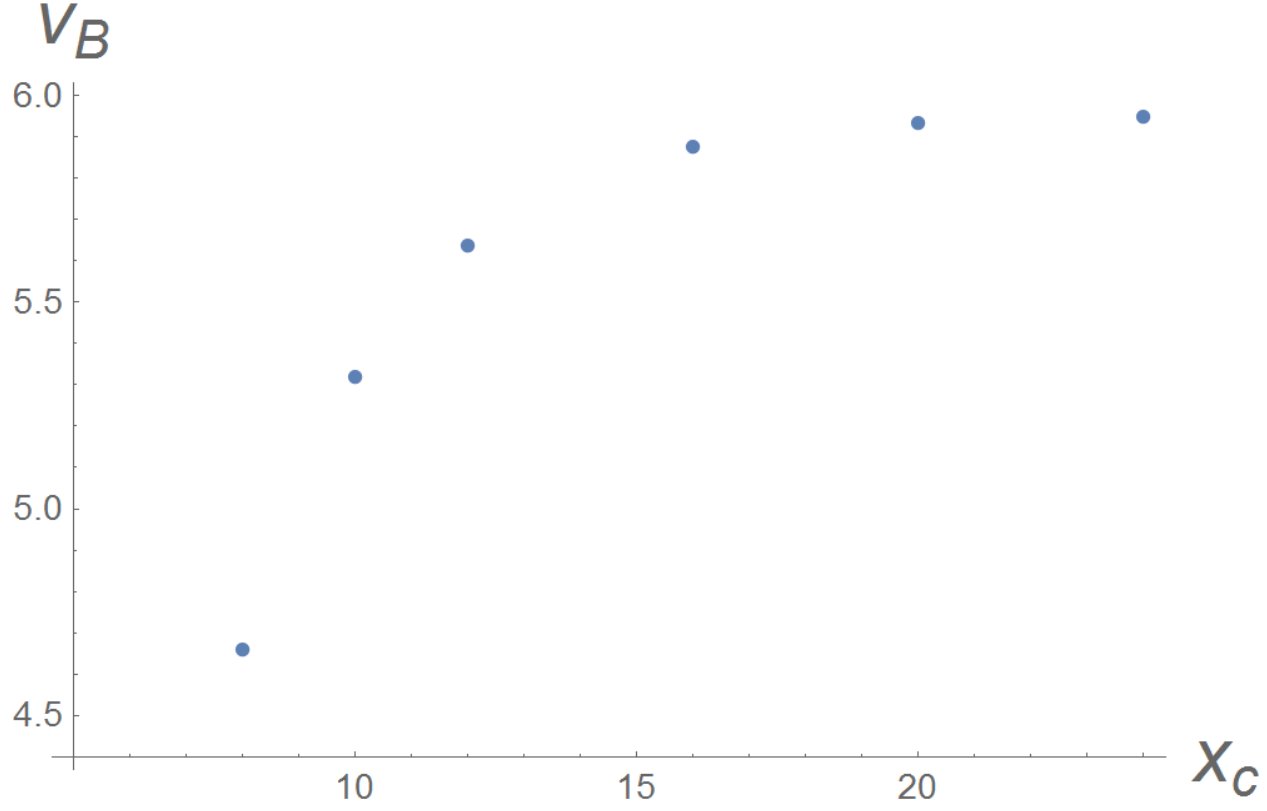}
\caption{The fitted butterfly velocity as a function of the center location $x_c$.}
\label{spatialbutterfly}
\end{figure}

To look at the spatial dependence of the butterfly velocity, we will study the OTOC $F_{x_c -a,x_c +a}(t)$ with different $a=0,1,2,...$ . We fit it with the form:
\begin{equation}
F_{x_c -a,x_c +a}(t)\propto \exp \left[ \lambda_L (t- \frac{2a}{v_B} )\right],
\end{equation}
where we assume that $\lambda_L$ and $v_B$ only depend on the center position $x_c$. In numerics, we can check that this assumption works well for small $a$. The dependence of $v_B$ on $x_c$ is shown in Fig. \ref{spatialbutterfly}.
From the plot we can see that,  the closer the center position $x_c$ is to the SYK$_{\psi}$ system, the smaller the butterfly velocity. This again shows that the presence of the larger system can also slow down the propagation of quantum chaos through space. The spatial dependence of Lyapunov exponent and butterfly velocity also backward prove the usefulness of the concepts in characterizing the degree of quantum chaos even locally.

\section{Summary and Outlook}

In this paper, we proposed several generalizations of the SYK models in which the characteristics of quantum chaos can be tuned by varying the coupling strength with a thermal bath. Owing to the separation of the scrambling time scales between the small and the large SYK system, and utilizing the nice large-$N$ structure of the SYK model, our set-up provides an exactly solvable model to illustrate the physics of how to tune the Lyapunov exponent and the butterfly velocity. We find a new temperature dependence law for the butterfly velocity as $v_B\propto T$ below the temperature scale induced by coupling to the bath. We also use simple numerics to illustrate the spatial dependence of the Lyapunov exponent and the butterfly velocity, which emphasizes the usefulness of the quantities as tools to diagnose the extent of quantum chaos locally. 

Since the Lyapunov exponent and the butterfly velocity are both tunable in our models, our work provides a broader platform to test various inequalities proposed for chaotic systems and to study the thermalization of chaotic systems. One possible generalization of our model is to impose an $U(1)$ charge symmetry, then one could study the thermoelectric transport properties of the system \cite{thermal transport} and see how they should be modified by interaction with the bath. 

Another interesting direction is to search for a holographic description of the system. Since the Schwarzian action of the original SYK model is shared by NAdS$_2$ gravity system, it is natural to ask whether the modified action given in this paper can be also found in some gravitational systems. More generally, it is still an open question whether the Lyapunov exponent in gravitational systems can be tuned by similar physical process inspired by this work. Similar questions can also be addressed on the behavior of butterfly velocity $v_B$. We defer these questions for further study.

\acknowledgments
We thank Wenbo Fu, Yingfei Gu, Chao-Ming Jian, Yi-Zhuang You and Xiao-Liang Qi  for helpful discussion. This work is supported by NSFC Grant No. 11325418 and MOST under Grant No. 2016YFA0301600.



\begin{thebibliography}{99}

\bibitem{Kitaev2}
A. Kitaev, \emph{talk given at KITP Program: Entanglement in Strongly-Correlated Quantum Matter}, 2015:\\
http://online.kitp.ucsb.edu/online/entangled15/kitaev/\\
http://online.kitp.ucsb.edu/online/entangled15/kitaev2/

\bibitem{SY}
S. Sachdev and J. Ye, \emph{Gapless spin-fluid ground state in a random quantum Heisenberg magnet}, \emph{Phys. Rev. Lett.} \textbf{70}, 3339 (1993).

\bibitem{Comments}
J. Maldacena and D. Stanford, \emph{Remarks on the Sachdev-Ye-Kitaev model}, \emph{Phys.Rev. D} {\bf 94} (2016) 106002.

\bibitem{spectrum1}
A. M. García-García and J. J. M. Verbaarschot, \emph{Spectral and thermodynamic properties of the Sachdev-Ye-Kitaev model}, \emph{Phys. Rev. D} \textbf{94}, 126010 (2016).

\bibitem{spectrum2}
Y. Liu, M. A. Nowak and I. Zahed, \emph{Disorder in the Sachdev-Yee-Kitaev Model}, \emph{arXiv}:1612.05233.

\bibitem{spectrum3}
A. M. García-García and J. J. M. Verbaarschot, \emph{Analytical Spectral Density of the Sachdev-Ye-Kitaev Model at finite N}, \emph{arXiv}:1701.06593.

\bibitem{Liouville}
D. Bagrets, A. Altland, and A. Kamenev, \emph{Sachdev-Ye-Kitaev model as Liouville quantum mechanics}, \emph{Nucl. Phys. B} \textbf{911} (2016) 191–205.

\bibitem{Liouville2}
D. Bagrets, A. Altland, A. Kamenev, \emph{Power-law out of time order correlation functions in the SYK model}, \emph{arXiv}:1702.08902.

\bibitem{SYK new}
E. Iyoda and T. Sagawa, \emph{Scrambling of Quantum Information in Quantum Many-Body Systems}, \emph{arXiv}:1704.04850.

\bibitem{SYK new2}
Thomas G. Mertens, Gustavo J. Turiaci and Herman L. Verlinde, \emph{Solving the Schwarzian via the Conformal Bootstrap}, \emph{1705.08408}.

\bibitem{SYK new3}
Razvan Gurau, \emph{The $i\varepsilon$ prescription in the SYK model}, \emph{1705.08581}.

\bibitem{bulk Yang}
J. Maldacena, D. Stanford and Z. Yang, \emph{Conformal symmetry and its breaking in two dimensional Nearly Anti-de-Sitter space},
\emph{Prog Theor Exp Phys} 2016 (12): 12C104.

\bibitem{bulk spectrum Polchinski}
J. Polchinski and V. Rosenhaus, \emph{The Spectrum in the Sachdev-Ye-Kitaev Model}, \emph{JHEP} 04 (2016) 001.

\bibitem{bulk2}
K. Jensen, \emph{Chaos in AdS 2 holography}, \emph{Phys. Rev. Lett.} \textbf{117}, 111601 (2016).

\bibitem{bulk3}
A. Jevicki and K. Suzuki, \emph{Bi-local holography in the SYK model: perturbations}, \emph{JHEP} 07 (2016) 007.

\bibitem{bulk4}
G. Mandal, P. Nayak, and S. R. Wadia, \emph{Virasoro coadjoint orbits of SYK/tensor-models and emergent two-dimensional quantum gravity}, \emph{arXiv}:1702.04266.

\bibitem{bulk5}
D. J. Gross and V. Rosenhaus, \emph{JHEP} 05 (2017) 092.

\bibitem{syk-bh}
J. S. Cotler, G. G.-Ari, M. Hanada, J. Polchinski, P. Saad, S. H. Shenker, D. Stanford, A. Streicher and M. Tezuka, \emph{Black Holes and Random Matrices}, \emph{arXiv}:1611.04650.

\bibitem{SYK g new1}
J. Maldacena, D. Stanford and Z. Yang, \emph{Diving into traversable wormholes}, \emph{arXiv}:1704.05333.

\bibitem{SYK g new2}
S. R. Das, A. Jevicki and K. Suzuki, \emph{Three Dimensional View of the SYK/AdS Duality}, \emph{arXiv}:1704.07208.

\bibitem{SYK g new3}
J. M. Magan, \emph{De Finetti theorems and entanglement in large-N theories and gravity}, \emph{arXiv}:1705.03048.

\bibitem{Kitaev1}
A. Kitaev,  \emph{talk given at Fundamental Physics Prize Symposium}, Nov.10, 2014:\\ http://online.kitp.ucsb.edu/online/joint98/kitaev/ 

\bibitem{bh1} 
S. H. Shenker and D. Stanford, \emph{Black holes and the butterfly effect}, \emph{JHEP} 03 (2014) 067.

\bibitem{bh2} 
S. H. Shenker and D. Stanford, \emph{Multiple shocks}, \emph{JHEP} 12 (2014) 046.

\bibitem{bh3} 
S. H. Shenker and D. Stanford, \emph{Stringy effects in scrambling}, \emph{JHEP} 05 (2015) 132.

\bibitem{otoc cft}
D. A. Roberts and D. Stanford, \emph{Diagnosing chaos using four-point functions in two-dimensional conformal field theory}, \emph{Phys. Rev, Lett.} \textbf{115}, 131603 (2015).

\bibitem{otoc anyon}
Y. Gu and X.-L. Qi, \emph{Fractional statistics and the butterfly effect}, \emph{JHEP} 08 (2016) 129.

\bibitem{OTOC-Ruihua MBL1}
R. Fan, P. Zhang, H. Shen and H. Zhai, \emph{Out-of-Time-Order Correlation for Many-Body Localization}, \emph{Science Bulletin}, 2017, 62(10): 707-711.

\bibitem{OTOC-MBL2}
X. Chen, T. Zhou, D. A. Huse and E. Fradkin, \emph{Out-of-time-order correlations in many-body localized and thermal phases}, \emph{Annalen der Physik}, 1521-3889, 1600332 2016.

\bibitem{OTOC-MBL3}
Y. Huang, Y.-L. Zhang, and X. Chen, \emph{Out-of-Time-Ordered Correlator in Many-Body Localized Systems}, \emph{Annalen der Physik}, 201600318.

\bibitem{OTOC-MBL4}
Y. Chen, \emph{Universal Logarithmic Scrambling in Many Body Localization},
\emph{arXiv}:1608.02765.

\bibitem{OTOC-MBL5}
 R.-Q. He and Z.-Y. Lu, \emph{Characterizing Many-Body Localization by Out-of-Time-Ordered Correlation}, \emph{Phys. Rev. B} 95, 054201.

\bibitem{OTOC-MBL6}
B. Swingle and D. Chowdhury, \emph{Slow scrambling in disordered quantum systems}, \emph{Phys. Rev. B} 95, 060201(R).

\bibitem{OTOC-Boson Hubbard}
H. Shen, P. Zhang, R. Fan and H. Zhai, \emph{Out-of-Time-Order Correlation at a Quantum Phase Transition}, \emph{arXiv}:1608.02438.

\bibitem{OTOC-Keldysh}
I. L. Aleinera, L. Faorob and L. B. Ioffe, \emph{Microscopic model of quantum butterfly effect: Out-of-time-order correlators and traveling combustion waves}, \emph{Annals of Physics}, Volume 375, December 2016, Pages 378–406.

\bibitem{OTOC-quantum}
K. Hashimoto, K. Murata and R. Yoshii, \emph{Out-of-time-order correlators in quantum mechanics}, \emph{arXiv}:1703.09435.

\bibitem{OTOC-classical}
J. S. Cotler, D. Ding and G. R. Penington, \emph{Out-of-time-order Operators and the Butterfly Effect}, \emph{arXiv}:1704.02979.

\bibitem{OTOC-Yao}
N. Y. Yao, F. Grusdt, B. Swingle, M. D. Lukin, D. M. Stamper-Kurn, J. E. Moore and E. A. Demler, \emph{Interferometric Approach to Probing Fast Scrambling}, \emph{arXiv}:1607.01801.

\bibitem{OTOC-quantum channel}
P. Hosur, X.-L. Qi, D. A. Roberts and B. Yoshida, \emph{Chaos in quantum channels}, \emph{JHEP} 02 (2016) 004.

\bibitem{OTOC-Luttinger}
B. Dóra and R. Moessner, \emph{Out-of-time-ordered density correlators in Luttinger liquids}, \emph{arXiv}:1612.00614.

\bibitem{OTOC-protocal}
B. Swingle, G. Bentsen, M. S.-Smith and P. Hayden, \emph{Measuring the scrambling of quantum information}, \emph{Phys. Rev. A} 94, 040302(R).

\bibitem{OTOC new}
P. Caputa, T. Numasawa and A. Veliz-Osorio, \emph{Out-of-time-ordered correlators and purity in rational conformal field theories}, \emph{Prog Theor Exp Phys} 2016 (11): 113B06.

\bibitem{OTOC-exp1}
M. Gärttner, J. G. Bohnet, A. Safavi-Naini, M. L. Wall, J. J. Bollinger and A. Rey, \emph{Measuring out-of-time-order correlations and multiple quantum spectra in a trapped ion quantum magnet}, \emph{Nature Physics}: 10.1038/nphys4119

\bibitem{OTOC-exp2}
J. Li, R. Fan, H. Wang, B. Ye, B. Zeng, H. Zhai, X. Peng and J. Du, \emph{Measuring out-of-time-order correlators on a nuclear magnetic resonance quantum simulator}, \emph{Phys. Rev. X}, to appear. \emph{arXiv}:1609.01246. 

\bibitem{prove} 
J. Maldacena, S. H. Shenker and D. Stanford, \emph{A bound on chaos}, \emph{JHEP} 08 (2016) 106.

\bibitem{numerics wenbo}
W. Fu and S. Sachdev, \emph{Numerical study of fermion and boson models with infinite-range random interactions}, \emph{Phys. Rev. B} \textbf{94}, 035135 (2016).

\bibitem{wenbo susy}
W. Fu, D. Gaiotto, J. Maldacena, and S. Sachdev, \emph{Supersymmetric Sachdev-Ye-Kitaev
	models}, \emph{Phys. Rev. D} \textbf{95} (2017) 026009.

\bibitem{susy2}
T. Li, J. Liu, Y. Xin and Y. Zhou, \emph{Supersymmetric SYK model and random matrix theory}, \emph{arXiv}:1702.01738.

\bibitem{susy3}
S. Förste, I. Golla, \emph{Nearly $AdS_2$ sugra and the super-Schwarzian}, \emph{Physics Letters B}.2017.05.039

\bibitem{Yingfei1}
Y. Gu, X.-L. Qi and D. Stanford, \emph{Local criticality, diffusion and chaos in generalized	Sachdev-Ye-Kitaev models},
\emph{arXiv}:1609.07832.

\bibitem{Yingfei2}
Y. Gu, A. Lucas and X.-L. Qi, \emph{Energy diffusion and the butterfly effect in inhomogeneous Sachdev-Ye-Kitaev chains},
\emph{arXiv}:1702.08462.
 
\bibitem{Altman}
S. Banerjee and E. Altman, \emph{Solvable model for a dynamical quantum phase transition from fast to slow scrambling}, \emph{Phys. Rev. B} 95, 134302.

\bibitem{sk jian}
S.-K. Jian and H. Yao, \emph{Solvable SYK models in higher dimensions: a new type of many-body localization transition}, \emph{arXiv}:1703.02051.


\bibitem{yyz condensation}
Z. Bi, C.-M. Jian, Y.-Z. You, K. A. Pawlak, and C. Xu, \emph{Instability of the non-Fermi liquid state of the Sachdev-Ye-Kitaev Model}, \emph{Phys. Rev. B} 95, 205105.

\bibitem{our}
X. Chen, R. Fan, Y. Chen, H. Zhai and P. Zhang, \emph{Competition between Chaotic and Non-Chaotic Phases in a Quadratically Coupled Sachdev-Ye-Kitaev Model}, \emph{arXiv}:1705.03406.

\bibitem{Balent}
X.-Y. Song, C.-M. Jian and L. Balents, \emph{A strongly correlated metal built from Sachdev-Ye-Kitaev models}, \emph{arXiv}:1705.00117.

\bibitem{thick}
D.V. Khveshchenko, \emph{Thickening and sickening the SYK model}, \emph{arXiv}:1705.03956.

\bibitem{generalization 1}
D. J. Gross and V. Rosenhaus, \emph{A Generalization of Sachdev-Ye-Kitaev},
\emph{arXiv}:1610.01569.

\bibitem{no disorder1}
B. Michel, J. Polchinski, V. Rosenhaus, and S. J. Suh, \emph{Four-point function in the IOP matrix model}, \emph{JHEP} \textbf{05} (2016) 048.

\bibitem{no disorder2}
E. Witten, \emph{An SYK-Like Model Without Disorder}, \emph{arXiv}:1610.09758.

\bibitem{no disorder3}
I. R. Klebanov and G. Tarnopolsky, \emph{Uncolored random tensors, melon diagrams, and the Sachdev-Ye-Kitaev models}, \emph{Phys. Rev. D} 95, 046004.

\bibitem{no disorder4}
C. Peng, M. Spradlin, and A. Volovich, \emph{A Supersymmetric SYK-like Tensor Model}, \emph{arXiv}:1612.03851.

\bibitem{no disorder5}
V. Bonzom, L. Lionni and A. Tanasa, \emph{Diagrammatics of a colored SYK model and of an SYK-like tensor model, leading and next-to-leading orders}, \emph{Journal of Mathematical Physics 58,} 052301 (2017).

\bibitem{no disorder6}
R. Gurau, \emph{The complete 1/N expansion of a SYK--like tensor model}, \emph{Nuclear Physics B}, Volume 916, March 2017, Pages 386–401.

\bibitem{no disorder7}
T. Nishinaka and S. Terashima, \emph{A Note on Sachdev-Ye-Kitaev Like Model without Random Coupling}, \emph{arXiv}:1611.10290.

\bibitem{no disorder8}
 C. Krishnan, S. Sanyal and P. N. Bala Subramanian, \emph{Quantum Chaos and Holographic Tensor Models}, \emph{JHEP} 1703, 056 (2017).
 
\bibitem{no disorder9}
R. Gurau, \emph{Quenched equals annealed at leading order in the colored SYK model}, \emph{arXiv}:1702.04228.

\bibitem{no disorder10}
C. Krishnan, K. V. P. Kumar and S. Sanyal, \emph{Random Matrices and Holographic Tensor Models}, \emph{arXiv}:1703.08155

\bibitem{no disorder new}
P. Narayan and J. Yoon, \emph{SYK-like Tensor Models on the Lattice}, \emph{arXiv}:1705.01554.

\bibitem{thermal transport}
R. A. Davison, W. Fu, A. Georges, Y. Gu, K. Jensen, and S. Sachdev, \emph{Thermoelectric transport in disordered metals without quasiparticles: the SYK models and holography}, \emph{Phys. Rev. B} 95, 155131.

\bibitem{high-D1}
M. Berkooz, P. Narayan, M. Rozali, and J. Simn, \emph{Higher Dimensional Generalizations of the SYK Model}, \emph{arXiv}:1610.02422.

\bibitem{high-D2-con}
G. Turiaci and H. Verlinde, \emph{Towards a 2d QFT Analog of the SYK Model},
\emph{arXiv}:1701.00528.

\bibitem{generalization 2}
C. Peng, \emph{Vector models and generalized SYK models}, \emph{arXiv}:1704.04223.

\bibitem{transition1}
C.-M. Jian, Z. Bi and C. Xu, \emph{A model for continuous thermal Metal to Insulator Transition}, \emph{arXiv}:1703.07793.

\bibitem{new gsyk1}
C. Peng, \emph{Vector models and generalized SYK models}, \emph{arXiv}:1704.04223.

\bibitem{vb T-1/3 energy}
A. A. Patel and S. Sachdev, \emph{Quantum chaos on a critical Fermi surface}, \emph{Proc. Nat. Acad. Sci.} \textbf{114} (2017) 1844–1849.

\bibitem{Blake}
M. Blake, \emph{Universal Charge Diffusion and the Butterfly Effect in Holographic Theories	
}, \emph{Phys. Rev. Lett.} \textbf{117}, 091601.

\bibitem{holographic review}
S. A. Hartnoll, A. Lucas and S. Sachdev, \emph{Holographic quantum matter}, \emph{arXiv}:1612.07324.

\bibitem{Sachdev new}
M. Blake, R. A. Davison and S. Sachdev, \emph{Thermal diffusivity and chaos in metals without quasiparticles}, \emph{arXiv}: arXiv:1705.07896.

\bibitem{vb bound}
S. A. Hartnoll, \emph{Theory of universal incoherent metallic
transport}, \emph{Nature Phys.} \textbf{11} (2015) 54.

\bibitem{vb energy1}
M. Baggioli, B. Gout\'{e}raux, E. Kiritsis, and W.-J. Li, \emph{Higher derivative corrections to incoherent metallic transport in holography}, \emph{arXiv}:1612.05500.

\bibitem{vb energy2}
M. Blake and A. Donos, \emph{Diffusion and Chaos from near AdS2 horizons}, \emph{arXiv}:1611.09380.

\bibitem{vbbh}
M. Blake, \emph{Universal diffusion in incoherent black holes}, \emph{Phys. Rev. D} 94, 086014.

\bibitem{Wenbo}
Wenbo Fu, priviate communication. 




\end{thebibliography}
\end{document}